\documentclass[conference, final]{IEEEtran}
\IEEEoverridecommandlockouts
\usepackage[rgb, table]{xcolor}
\usepackage{pgfplots}
\usepackage{pgfkeys}
    \def\addlegendimage{\csname pgfplots@addlegendimage\endcsname}
\pgfplotsset{compat=1.8}
\usepgfplotslibrary{statistics}
\usepackage{subcaption}
\usepackage{amsmath}
\usepackage{etoolbox}
\usepackage{amssymb}
\usepackage{pifont}
\usepackage{enumitem}
\usepackage{setspace}
\usepackage[margin=1in]{geometry} 
\usepackage{xcolor}
\usepackage{amsthm}
\usepackage{mathtools}
\usepackage{braket}
\usepackage{soul}
\usepackage{algorithm}
\usepackage[noend]{algpseudocode}
\usepackage[format=hang,font=small,labelfont=bf]{caption}

\usepackage{mdframed}
\usepackage{threeparttable}

\usepackage{multirow}

\usepackage{tikz}
\usetikzlibrary{plotmarks}
\usepackage{pgfplots}
\usepackage[normalem]{ulem}
\usepackage[numbers,sort&compress]{natbib}
\usetikzlibrary{patterns}
\pgfplotsset{compat=1.12}
\usetikzlibrary{backgrounds}
\pgfplotsset{major grid style={dotted}}
\usetikzlibrary{arrows, arrows.meta}

\usepackage{orcidlink}
\definecolor{store_color}{HTML}{375e97}
\definecolor{assign_color}{HTML}{fb6542}
\definecolor{share_color}{HTML}{ffbb00}
\definecolor{access_color}{HTML}{3f681c}
\definecolor{write_color}{HTML}{84817a}
\definecolor{cal_read}{HTML}{0000ff}
\definecolor{cal_write}{HTML}{ff0000}
\definecolor{mygreen}{HTML}{bef6d0}
\definecolor{myred}{HTML}{ffc6c6}
\definecolor{myorange}{HTML}{ffe2c3}

\newcommand{\storeshape}{o}
\newcommand{\assignshape}{*}
\newcommand{\shareshape}{s}
\newcommand{\accessshape}{diamond}
\newcommand{\writeshape}{square}
\newcommand{\calreadshape}{diamond}
\newcommand{\calwriteshape}{square}

\newcommand{\RgeG}{\mathsf{RgeG}}
\newcommand{\MemG}{\mathsf{MemG}}
\newcommand{\HshG}{\mathsf{HshG}}
\newcommand{\PedG}{\mathsf{PedG}}
\newcommand{\secparam}{\lambda}
\newcommand{\pp}{\mathsf{pp}}
\newcommand{\fid}{\filehash}
\newcommand{\assign}{\mathsf{own}}

\newcommand{\access}{\mathsf{acc}}

\newcommand{\share}{\mathsf{shr}}
\newcommand{\negl}{\mathsf{negl}}

\newcommand{\sn}{\mathsf{sn}}
\newcommand{\cm}{\mathsf{cm}}

\newcommand{\curr}{\mathsf{curr}}

\newcommand{\mypath}{\mathsf{path}}
\newcommand{\own}{\mathsf{own}}

\newcommand{\tx}{\mathsf{tx}}
\newcommand{\addr}{\mathsf{addr}}

\newcommand{\s}{s}

\newcommand{\m}{m}

\newcommand{\rt}{\mathsf{rt}}
\newcommand{\ts}{\mathsf{ts}}

\newcommand{\cipher}{c}
\newcommand{\x}{x}
\newcommand{\w}{w}

\newcommand{\h}{h}

\newcommand{\file}{D}

\newcommand{\pk}{\mathsf{pk}}
\newcommand{\sigsk}{\ensuremath{\mathsf{sk}_{\mathsf{sig}}}}
\newcommand{\sigpk}{\ensuremath{\mathsf{pk}_{\mathsf{sig}}}}

\newcommand{\sig}{\mathsf{sig}}

\newcommand{\store}{\mathsf{str}}
\newcommand{\filehash}{v}

\newcommand{\zkp}{\ensuremath{\mathsf{zkp}}}

\newcommand{\addsk}{\ensuremath{\mathsf{sk}_{\mathsf{adr}}}}
\newcommand{\addpk}{\ensuremath{\mathsf{pk}_{\mathsf{adr}}}}

\newcommand*{\Scale}[2][4]{\scalebox{#1}{$#2$}}%

\newcommand{\Enc}{\mathsf{Enc}}
\newcommand{\Dec}{\mathsf{Dec}}
\newcommand{\Sig}{\mathsf{Sign}}
\newcommand{\zkVerify}{\ensuremath{\zkp.\mathcal{V}}}
\newcommand{\Verify}{\mathsf{Verify}}
\newcommand{\zkProve}{\ensuremath{\zkp.\mathcal{P}}}
\newcommand{\VerifyLog}{\mathsf{VerifyRecord}}
\newcommand{\Com}{\mathsf{Com}}
\newcommand{\Init}{\mathsf{Init}}

\newcommand{\PRF}{\mathsf{PRF}}

\newcommand{\CRH}{H}

\newcommand{\zkKeyGen}{\ensuremath{\zkp.\mathcal{G}}}

\newcommand{\zkSim}{\ensuremath{S}}
\newcommand{\Sim}{\ensuremath{S}}
\newcommand{\keygen}{\mathsf{Gen}}
\newcommand{\Setup}{\mathsf{Com}.\mathcal{G}}

\newcommand{\AssignOwner}{\mathsf{AssignOwner}}

\newcommand{\Store}{\mathsf{Store}}
\newcommand{\CreateAddress}{\mathsf{Register}}
\newcommand{\Share}{\mathsf{Share}}
\newcommand{\Access}{\mathsf{Access}}

\renewcommand{\Pr}{\mathsf{Pr}}

\newcommand{\owner}{\mathcal{O}}
\newcommand{\user}{\mathcal{U}}
\newcommand{\adv}{\mathcal{A}}

\newcommand{\chal}{\mathcal{C}}
\newcommand{\provider}{\mathcal{S}}

\newcommand{\ledger}{\mathcal{L}} 

\newcommand{\yesbw}{\checkmark}
\newcommand{\nobw}{\xmark}
\newcommand{\tokenstore}{\ensuremath{\mathsf{tk}_{\store}}}
\newcommand{\tokenowner}{\ensuremath{\mathsf{tk}_{\mathsf{own}}}}
\newcommand{\tokenshare}{\ensuremath{\mathsf{tk}_{\share}}}
\newcommand{\tokenaccess}{\ensuremath{\mathsf{tk}_{\access}}}
\newcommand{\token}{\ensuremath{\mathsf{tk}}}

\newcommand{\encryption}{\mathcal{E}}
\newcommand{\signature}{\Sigma}

\newcommand{\trnm}{\normalfont \texttt{NMGame}}
\newcommand{\lind}{\ensuremath{\normalfont \texttt{IndGame}}}

\newcommand{\sys}{\ensuremath{\mathsf{Harpocrates}}}

\newcommand{\xmark}{\ding{55}}

\newcommand{\cmark}{\ding{51}}

\newcommand{\sample}{\xleftarrow{\$}}

\newcommand{\FuncHeader}[1]{\underline{#1}\vspace{1mm}}

\makeatletter
\newcommand{\comment}[1]{\Statex \hskip\ALG@thistlm \(\triangleright\) \textit{#1}}
\makeatother
\makeatletter
\newcommand{\commentx}[1]{ \hskip\ALG@thistlm \(\triangleright\) \textit{#1}}
\makeatother

\renewcommand{\set}[1]{\ensuremath{\mathcal{#1}}}

\newcommand{\DSAL}{PIAL}

\usetikzlibrary{patterns}

\newlength{\hatchspread}
\newlength{\hatchthickness}
\newlength{\hatchshift}

\tikzset{hatchspread/.code={\setlength{\hatchspread}{#1}},
         hatchthickness/.code={\setlength{\hatchthickness}{#1}},
         hatchshift/.code={\setlength{\hatchshift}{#1}},
         hatchcolor/.code={\renewcommand{\hatchcolor}{#1}}}
\tikzset{hatchspread=1.8pt,
         hatchthickness=0.4pt,
         hatchshift=0pt,
         }
\pgfdeclarepatternformonly[\hatchspread,\hatchthickness,\hatchshift]
   {custom north west lines}
   {\pgfqpoint{\dimexpr-2\hatchthickness}{\dimexpr-2\hatchthickness}}
   {\pgfqpoint{\dimexpr\hatchspread+2\hatchthickness}{\dimexpr\hatchspread+2\hatchthickness}}
   {\pgfqpoint{\dimexpr\hatchspread}{\dimexpr\hatchspread}}
   {
    \pgfsetlinewidth{\hatchthickness}
    \pgfpathmoveto{\pgfqpoint{0pt}{\dimexpr\hatchspread+\hatchshift}}
    \pgfpathlineto{\pgfqpoint{\dimexpr\hatchspread+0.15pt+\hatchshift}{-0.15pt}}
    \ifdim \hatchshift > 0pt
      \pgfpathmoveto{\pgfqpoint{0pt}{\hatchshift}}
      \pgfpathlineto{\pgfqpoint{\dimexpr0.15pt+\hatchshift}{-0.15pt}}
    \fi
    \pgfsetstrokecolor{access_color}
    \pgfusepath{stroke}
   }

\pgfdeclarepatternformonly{crosshatchh}{\pgfqpoint{-1pt}{-1pt}}{\pgfqpoint{4pt}{4pt}}{\pgfqpoint{2pt}{2pt}}%
{%
  \pgfsetlinewidth{0.4pt}%
  \pgfpathmoveto{\pgfqpoint{2.1pt}{0pt}}%
  \pgfpathlineto{\pgfqpoint{0pt}{2.1pt}}%
  \pgfpathmoveto{\pgfqpoint{0pt}{0pt}}%
  \pgfpathlineto{\pgfqpoint{2.1pt}{2.1pt}}%
  \pgfusepath{stroke}%
}%

\pgfdeclarepatternformonly[\hatchspread,\hatchthickness,\hatchshift]
   {custom north east lines}
   {\pgfqpoint{\dimexpr-2\hatchthickness}{\dimexpr-2\hatchthickness}}
   {\pgfqpoint{\dimexpr\hatchspread+2\hatchthickness}{\dimexpr\hatchspread+2\hatchthickness}}
   {\pgfqpoint{\dimexpr\hatchspread}{\dimexpr\hatchspread}}
   {
    \pgfsetlinewidth{\hatchthickness}
    \pgfpathmoveto{\pgfqpoint{0pt}{0pt}}
    \pgfpathlineto{\pgfqpoint{\dimexpr\hatchspread+0.15pt+\hatchshift}{\dimexpr\hatchspread+0.15pt+\hatchshift}}
    \ifdim \hatchshift > 0pt
      \pgfpathmoveto{\pgfqpoint{0pt}{0pt}}
      \pgfpathlineto{\pgfqpoint{\dimexpr0.15pt+\hatchshift}{\dimexpr0.15pt+\hatchshift}}
    \fi
    \pgfsetstrokecolor{assign_color}
    \pgfusepath{stroke}
   }

\setlist[itemize]{leftmargin=.15in, topsep=1pt, itemsep = .3em}
\newcommand{\algrule}[1][.2pt]{\par\vskip.\baselineskip\hrule height #1\par\vskip.5\baselineskip}

\theoremstyle{definition}

\newcommand{\yes}{{\color{green}\cmark}}
\newcommand{\no}{{\color{red}\xmark}}

\makeatletter
\def\thmheadbrackets#1#2#3{%
	\thmname{#1}\thmnumber{\@ifnotempty{#1}{ }\@upn{#2}}%
	\thmnote{ {\the\thm@notefont[#3]}}}
\makeatother

\newtheoremstyle{myDefinitionStyle}
{}
{}
{}
{}
{%
	\bfseries
}
{.}
{ }
{\thmname{#1}\thmnumber{ #2}\thmnote{ (#3)}}%

\newcommand{\PP}[1]{
	\vspace{2px}
	\noindent{\bf {#1}{.}}
}
\newcommand{\PC}[1]{
	\vspace{2px}
	\noindent{\bf \IfEndWith{#1}{:}{#1}{#1:}}
}

\newcommand{\hs}{\hspace{1.5em}}

\newtheoremstyle{myTheoremStyle}
{}
{}
{\itshape}
{}
{%
	\bfseries
}
{.}
{ }
{\thmname{#1}\thmnumber{ #2}\thmnote{ (#3)}}%

\theoremstyle{myDefinitionStyle}
\newtheorem{Definition}{Definition}

\theoremstyle{myTheoremStyle}
\newtheorem{Theorem}{Theorem}

\usepackage{amsthm}

\usepackage{hyperref}

\def\Snospace~{\S{}}

\hypersetup{breaklinks,colorlinks,citecolor=blue,linkcolor=blue}

\def\BibTeX{{\rm B\kern-.05em{\sc i\kern-.025em b}\kern-.08em
    T\kern-.1667em\lower.7ex\hbox{E}\kern-.125emX}}

\begin{document}

\title{\sys: Privacy-Preserving and Immutable Audit Log for Sensitive Data Operations
}
\author{
\IEEEauthorblockN{Mohit Bhasi Thazhath, Jan Michalak, Thang Hoang}
\IEEEauthorblockA{\textit{Department of Computer Science, Virginia Tech, USA} \\
\{mohitbt, janmichalak, thanghoang\}@vt.edu}
}

\newcommand{\thang}[1]{{\small{\color{blue}thang:#1}}}

\maketitle
\begin{abstract}
The audit log is a crucial component to monitor fine-grained operations over sensitive data (e.g., personal, health) for security inspection and assurance.
Since such data operations can be highly sensitive, it is vital to ensure that the audit log achieves not only validity and immutability, but also confidentiality against active threats to standard data regulations (e.g., HIPAA) compliance.
Despite its critical needs, state-of-the-art privacy-preserving audit log schemes (e.g., Ghostor (NSDI '20), Calypso (VLDB '19)) do not fully obtain a high level of privacy, integrity, and immutability simultaneously, in which certain information (e.g., user identities) is still leaked in the log.
In this paper, we propose \sys, a new privacy-preserving and immutable audit log scheme.
\sys~permits data store, share, and access operations to be recorded in the audit log without leaking sensitive information (e.g., data identifier, user identity), while permitting the validity of data operations to be publicly verifiable.
\sys~makes use of blockchain techniques to achieve immutability and avoid a single point of failure, while cryptographic zero-knowledge proofs are harnessed for confidentiality and public verifiability. 
We analyze the security of our proposed technique and prove that it achieves non-malleability and indistinguishability.
We fully implemented \sys~and evaluated its performance on a real blockchain system (i.e., Hyperledger Fabric) deployed on a commodity platform (i.e., Amazon EC2).
Experimental results demonstrated that \sys~is highly scalable and achieves practical performance.

%
%
%
%

\end{abstract}
\thispagestyle{plain}
\pagestyle{plain}
   	\section{Introduction} \label{sec.introduction}

Remote data storage systems have become predominant in the past decade due to the growth of cloud computing.
Cloud Service Providers (CSPs) have dedicated resources to store data, and have developed a suite of services to support it such as AWS S3, Azure Blobs, GCP Buckets.
The widespread use of cloud data storage has made it easy to share/access remote data. 

Given that CSPs generally maintain a vast amount of user data, 
many of which can be highly sensitive (e.g., personal, health), 
it is vital to maintain an audit log that records all the data operations (e.g., sharing, access activities) 
to support system integrity inspection and critical security assurances such as intrusion detection and problem analysis \cite{tifs}. 
Many data audit log schemes have been proposed to record fine-grained operations (e.g., read, write, share) over sensitive data (e.g., Electronic Health Records (EHRs) \cite{Liu_Access, Medrec}, supply chains \cite{Baralla,agri_suuply,poms,iot_supply}) in a tamper-resistant and immutable manner.
%
%
%

Despite their merits, achieving the immutability and integrity of the audit log  may not be sufficient to ensure security against active threats.
This is because the audit log contains all data operations (i.e., metadata), which can reveal significant sensitive information.
By analyzing the data log, the adversary can obtain the needed information  without having to access the actual data \cite{MetadataLeakage:zhou2020mobilogleak}.
Such metadata leakage has become notorious in related areas such as communication surveillance. 
A former NSA General Counsel stated, ``\emph{Metadata absolutely tells you everything about somebody's life}'' \cite{NYRB13:Alan-Rusbridger-Metadata}.
Due to the sensitivity of metadata, it is vital to ensure integrity and confidentiality of audit logs as indicated in standard data regulations such as The Health Insurance Portability and Accountability Act of 1996 (HIPAA) \cite{hipaa}.

To enable both privacy and integrity, several privacy-preserving data audit log schemes have been proposed.
Preliminary constructions permit data operations to be logged in a single server with privacy and integrity guarantees using cryptographic tools such as digital signatures and symmetric encryption \cite{jsyst, ma}.
Despite their merits, such centralized approaches suffer from a single point of failure, in which the corrupted server can compromise the validity and confidentiality of the audit log. 
To address single point of failure, several decentralized privacy-preserving audit log  approaches have been proposed using blockchain techniques \cite{MeDShare,BPDS_globecom2018,pp_sc_1,pp_sc_2, droplet, ghostor, calypso_vldb20}.
However, there are certain limitations to these techniques.
For example, Droplet \cite{droplet} offers confidentiality but not anonymity.
Ghostor \cite{ghostor} records data sharing activities with anonymity, but the validity can only be verified privately by the data owner.
This requires {all} the users to participate in the audit process to verify the validity of the data operations, thereby reducing the audit transparency. 
Calypso \cite{calypso_vldb20} achieves partial anonymity (i.e., leaks data owner identity) and private verifiability.

%
%

%

%






\PP{Research gaps}
Given that existing data logging techniques lack a certain degree of anonymity, integrity, confidentiality, and transparency, our objective is to design a new privacy-preserving audit log scheme that can offer all desirable security properties for standard data regulations compliance.


%

\PP{Contribution} 
In this paper, we propose \sys, a new privacy-preserving and immutable audit log scheme for sensitive data operations. 
\sys~permits data operations (i.e., store, share, access) to be recorded in the audit log with validity, confidentiality, and public verifiability guarantees.
%
%
\sys~makes use of blockchain technologies (e.g., distributed ledger) to achieve immutability and validity, while confidentiality and public verifiability are achieved using advanced cryptographic techniques such as zero-knowledge proofs \cite{bulletproof}.

\begin{table*}[ht!]
    \footnotesize
	\centering
	\begin{threeparttable}
		\caption{Comparison of \sys~with prior works.}
		
		\begin{tabular}{|c|c|c|c|c|c|c|}
			\hline
			\multirow{2}{0.1\textwidth}{\centering \textbf{Scheme}} &	\multirow{2}{0.13\textwidth}{\centering \textbf{Audit Log \\ Confidentiality}} &	\multirow{2}{0.13\textwidth}{\centering \textbf{Audit Log \\ Anonymity}} &	\multirow{2}{0.13\textwidth}{\centering \textbf{Temporal Access Control}} & \multicolumn{3}{c|}{\textbf{Validity Verifiability}} \\ 
\cline{5-7}
& & & &\textbf{Store Record} & \textbf{Share Record} & \textbf{Access Record} \\ \hline
	
			 Calypso \cite{calypso_vldb20} & \cellcolor{mygreen}  \yesbw & \cellcolor{myorange} Partial\textsuperscript{$\dagger$} & \cellcolor{myred}  \nobw & \cellcolor{myorange} Private & \cellcolor{myorange} Private & \cellcolor{myorange} Private \\
			 \hline
			 Ghostor \cite{ghostor} & \cellcolor{mygreen} \yesbw & \cellcolor{mygreen} \yesbw & \cellcolor{myred} \nobw & \cellcolor{myorange} Private & \cellcolor{myorange} Private & \cellcolor{myorange} Private \\
			 \hline
			 Droplet \cite{droplet} & \cellcolor{mygreen} \yesbw & \cellcolor{myred} \nobw\textsuperscript{$\dagger$} & \cellcolor{myred} \nobw & \cellcolor{myred} Not Verifiable & \cellcolor{mygreen} Public & \cellcolor{myred} Not Verifiable \\
			 \hline
			 \sys & \cellcolor{mygreen} \yesbw & \cellcolor{mygreen} \yesbw & \cellcolor{mygreen} \yesbw & \cellcolor{mygreen} Public & \cellcolor{mygreen} Public & \cellcolor{mygreen} Public \\ \hline			 
		\end{tabular}\label{lit_review}
		
		\begin{tablenotes}[flushleft]
			\item $\dagger$~Calypso and Droplet do not hide data owner identity.
			
		\end{tablenotes}
	\end{threeparttable}
	
\vspace{-6mm} 
\end{table*}

\sys~achieves the following properties.
\begin{itemize}
    \item \underline{\textit{Record Immutability}}: All the data operation records cannot be modified by anyone.
    \item \underline{\textit{Full Anonymity}}: The identity of the  data and the user(s) remain hidden all the time. This security guarantee is stronger than existing works  (e.g., \cite{calypso_vldb20, droplet}).
    \item \underline{\textit{Publicly Verifiable Validity}}: Validity of the data operations (e.g., whether the share/access is performed by an authorized user) can be publicly verified by anyone. This improves audit transparency such that all the data users do not need to participate during the audit process.
    
    \item \underline{\textit{Temporal Access Control}}: Restricts the time for which the data can be accessed. This is useful when operating on highly sensitive data like EHRs.
\end{itemize}

\autoref{lit_review} compares \sys~with state-of-the-art techniques. 
We rigorously analyze the security of our proposed technique and prove that it achieves standard security notions (non-malleability, indistinguishability). 
Finally, we implemented our technique and deployed it on a commodity platform (Amazon EC2) to evaluate its efficiency. Experimental results showed that \sys~is highly scalable (\autoref{sec.experiments}) and practical. Therefore, \sys~ can be used for sensitive data storage applications that require privacy-preserving audit log such as EHRs \cite{MeDShare,BPDS_globecom2018} or supply chains \cite{pp_sc_1, pp_sc_2}.

 \label{ch:introduction}
   	\section{Preliminaries}
\subsection{Notation}
We denote $||$ as the concatenation operator.
Let $\secparam$ be a security parameter, $\negl(\cdot)$ be a negligible function.
We denote $H : \{0,1\}^* \rightarrow \{0,1\}^\secparam $ as a collision-resistant hash function.
$r \gets \PRF(\s)$ is a pseudorandom function that outputs a pseudorandom $r$ given a seed $s$. 
Let $\encryption = (\keygen, \Enc, \Dec)$  be an asymmetric encryption, in which $(\addpk, \addsk) \gets \encryption.\keygen(1^\secparam)$ generates a public and private key pair given a security parameter $\secparam$; $\cipher \gets \encryption.\Enc({\addpk},\m)$ encrypts a plaintext $m$ under public key $\addpk$; $\m \gets \encryption.\Dec({\addsk},\cipher)$ decrypts a ciphertext $\cipher$ with private key $ \addsk $. 
Let $\signature = (\keygen, \Sig, \Verify)$ be a digital signature, where $(\sigpk, \sigsk) \gets \signature.\keygen(1^\secparam)$ generates a public and private key pair; $\sigma \gets \signature.\Sig({\sigsk}, \m)$ produces a signature $ \sigma $ for message $\m$ under private key $\sigsk$; $\{0,1\} \gets \signature.\Verify(\sigpk, m, \sigma)$ verifies whether $ \sigma $ is a valid signature of $ m $ using public key $\sigpk$.

\subsection{Cryptographic Building blocks}

\PP{Commitment Scheme} 
A commitment scheme is a tuple of Probabilistic Polynomial Time (PPT) algorithms $(\Setup,\Com)$ --
 
	\begin{itemize}[leftmargin=.18in, topsep=1pt, itemsep = .3em]
	\item $\pp \gets \Setup(1^\secparam)$: Given a security parameter $\secparam$, it outputs public parameters $\pp$. 
	
	\item $\cm \gets \Com_r(\m,\pp)$: Given a message $\m$ and a trapdoor $ r $, it outputs a commitment $\cm$. 
	
	
\end{itemize}

	
A commitment scheme is \textit{binding} if for any PPT adversary $\adv$ such that  $\pp \gets \Setup(1^\secparam)$, $(\m_0, \m_1, r_0, r_1) \gets \adv(\pp)$, it holds that $\Pr \left[\Com_{r_0}(m_0,\pp) = \Com_{r_1}(m_1,\pp) \land m_0 \ne m_1\right] \leq \negl(\secparam)$; 
\textit{hiding} if for any PPT $\adv$ such that $(m_0, m_1) \gets \adv(\pp), b \sample {0,1}, r \sample \{0,1\}^\secparam, \cm \gets \Com_{r_b}(m_b,\pp), b' \gets \adv(\cm)$, it holds that $|\Pr[b = b'] - \frac{1}{2}| \leq \negl(\lambda)$.

\PP{Merkle Tree} Merkle tree permits to commit to a set of values with the ability to reveal an element in the committed set later (proof of membership).
Given $\vec{\m} = (\m_1, \m_2, \m_3, \dotsc, \m_n)$,
it first initializes a binary tree $ T $ with $ n $ leaves and $ \CRH(m_i) $ be the $ i $-th leaf node. 
$ T $ is built bottom-up, where
each non-leaf node $ t $ of $ T $ is computed by computing the hash of its child nodes $ t_{\mathsf{l}} $ and $ t_{\mathsf{r}} $, i.e., $ t =  \CRH(t_{\mathsf{l}}||t_{\mathsf{r}})$.
To prove  $ m_i \in \vec{m}$, the prover reveals $ \mypath $ from $\rt$ to $\CRH(m_i)$, where $\rt$ is root of $T$ and $\mypath$ is the set of sibling nodes along the path from $m_i$ to $\rt$.
The verifier recomputes $ \rt $ using $\mypath$ for membership verification.

Given an updated set $ \vec{m'} $, 
where $ m_i \in \vec{m'} $ is the updated value.
The prover updates $i^{th}$ leaf with $H(m'_i)$, recomputes non-leaf nodes along the path from $\CRH(m'_i)$ to $ \rt $ in logarithmic time, and outputs updated root $ \rt' $.
%

\PP{Argument of Knowledge} 
An argument of knowledge for an NP relation $\mathcal{R}$ is a protocol between a prover $\mathcal{P}$ and a verifier $ \mathcal{V} $, in which $\mathcal{P}$ convinces $\mathcal{V}$ that it knows a witness $ w $ for some input in an NP language $ x \in \mathcal{L}$ such that $(x,w) \in \mathcal{R}$.
Formally speaking, a zero-knowledge argument of knowledge is a tuple of PPT algorithms $\mathsf{zkp}=(\mathcal{G},\mathcal{P},\mathcal{V})$ as follows.

	\begin{itemize}[leftmargin=.18in, topsep=1pt, itemsep = .3em]
        \item $\pp \gets \zkKeyGen(1^\secparam)$: Given a security parameter $\secparam$, it outputs public parameters $ \pp $.
        \item $\pi \gets \zkProve(\x, \w, \pp)$: Given a statement $x$ and  a witness $w$, it generates a proof $\pi$ to show $(\x,\w) \in \mathcal{R}$. 
        \item $\{0,1\} \gets \zkVerify(\x, \pi, \pp)$: Given a statement $x$ and a proof $\pi$, it outputs $1$ if $\pi$ is valid and $(x,w) \in \mathcal{R}$, else it outputs 0. 
    \end{itemize}

	An argument of knowledge is \textit{complete} if for any $(\x, \w) \in \mathcal{R}$, $ \pp \gets \zkKeyGen(1^\secparam)$,  $\pi \gets \zkProve(x, w,\pp)$, it holds that $\Pr \left[ \zkVerify(\x,\pi,\pp) = 1\right ] = 1$; is \textit{sound} iff for any PPT prover $ \mathcal{P}^* $, there exists a PPT extractor $ \mathcal{E} $, when given the entire execution and randomness of $ \mathcal{P}^* $, $ \mathcal{E} $~can extract a witness $ w $ such that $\pp \gets \zkKeyGen(1^\secparam)$, $\pi^* \gets \mathcal{P}^*(x,\pp)$, $ w \gets \mathcal{E}^{\mathcal{P}^*}(x,\pi^*,\pp)$ and
	$\Pr \left[\zkVerify(\x, \pi,\pp) = 1 \land (x,w) \notin \mathcal{R}\right] \leq\negl(\secparam)$; is \textit{zero-knowledge} iff for any PPT verifier $\mathcal{V}^*$ there exists simulators $(\zkSim_1,\zkSim_2)$ --
	
	\[\Scale[0.85]{
	\Pr \left[
	\begin{array}{r}
		\pp \gets \zkKeyGen(1^\secparam) \\
		\pi \gets \zkProve(\x, \w,\pp)\\
	    \zkVerify(x,\pi,\pp)= 1 
	\end{array}
	\right] 
	\approx
	\Pr \left[
	\begin{array}{r}
		\pp \gets \zkSim_1(1^\secparam) \\
		\pi \gets \zkSim_2(\x,\pp)\\
		\mathcal{V}^*(x,\pi,\pp) = 1
	\end{array}
	\right]}
	\]	
 \label{ch:prelim}
   	\section{Models}
\PP{System Model} In our system, there are three types of participants:
(i)~The data owner $\owner$ who owns some data; 
(ii)~The data user $ \user $ who would like to access $\owner$'s data;
(iii)~The storage provider $ \provider $ that provides storage facilities to store $\owner$'s data remotely.
Our system permits data storage, sharing, and access between these participants while, at the same time, \emph{recording} all the data operations in a distributed audit log for auditability.
%
A Privacy-preserving and Immutable Audit Log (\DSAL) scheme consists of the following PPT algorithms --

\begin{itemize}[leftmargin=.15in, topsep=1pt, itemsep = .3em]
	\item $ (\ledger,\pp) \gets \Init(1^\secparam, N)$: Given a security parameter $\lambda$, and the maximum number of supported operations $N$, it outputs an initial distributed audit log $\ledger$ and public parameters $ \pp $.
	
	\item $(\addpk,\addsk) \gets \CreateAddress(1^\secparam)$: Given a security parameter $ \secparam $, it outputs an address-key pair $(\addpk,\addsk)$ as an identifier for the participant. 
	
	\item $(\tokenstore,\tx_{\store}) \gets \Store(\file, \addsk^{\owner}, \addpk^{\provider}, \pp $): Given data $\file$, $\owner$'s private identity $ \addsk^{\owner} $,  $\provider$'s public identity $\addpk^{\provider}$, it outputs a store token $ \tokenstore $ and a store record $\tx_{\store} $.

	\item $(\tokenowner,\tx_{\assign}) \gets \AssignOwner( \tokenstore,\addsk^{\provider},\addpk^{\owner}, \pp)$: 
	Given $ \tokenstore$, $\provider$'s private identity $\addsk^{\provider}$, and $\owner$'s public identity $\addpk^{\owner}$, it outputs an ownership token $ \tokenowner $, which permits $\owner$ to share her data stored on $\provider$ later, and create a record of ownership  $\tx_{\assign}$.
	
	\item $(\tokenshare,\tx_{\share}) \gets \Share(\tokenowner,\addsk^{\owner}, \addpk^{\user},\ts, \pp)$: Given $ \tokenowner $, $\owner$'s private identity $\addsk^{\owner}$, $\user$'s public identity $\addpk^{\user}$, a share expiry timestamp $ \ts>0$, it outputs a data share token  $ \tokenshare $ that is only valid up to $ \ts $, and a record of share  $\tx_{\share}$. 
	
	\item $(\tokenaccess,\tx_{\access}) \gets \Access(\tokenshare,\addsk^{\user},\addpk^{\provider}. \pp)$: Given $ \tokenshare $, $\user$'s private identity $\addsk^{\user}$, $\provider$'s public identity $ \addpk^{\provider} $, it outputs a data access token $\tokenaccess$ (if $ \tokenshare $ is not expired) and a record of access $\tx_{\access}$.
	

	\item $ \{0,1\}\gets \mathsf{VerifyLog}(\tx,\set{L},\pp) $: Given a record $ \tx $ and audit log $\ledger$, it outputs $ 1 $ if the record is valid else $ 0 $.

\end{itemize} 

\PP{Threat and Security Models}
We assume participants do not trust each other, i.e., they can behave maliciously toward the records created by other participants. 
The adversary is curious about the record content, e.g., who is the owner of a particular data, whom the data is shared with or accessed by, and what data is being shared or accessed.
The adversary may also attempt to modify the record to compromise validity or to gain unauthorized access.
We aim to achieve integrity and confidentiality properties for the audit log via \textit{record non-malleability} and \textit{audit log indistinguishability} as follows.

\begin{Definition}[Record Non-Malleability] \label{lnm}
	Let $ \Pi $ be a candidate \DSAL~scheme. Consider experiment $\trnm_{\Pi, \adv}(\secparam)$ between the adversary $ \adv $ and the challenger $\chal$. $\chal$ samples $ \pp \gets \Init(1^\secparam) $ and sends $ \pp $ to $ \adv $. Then, $\chal$ initializes a \DSAL~oracle $\mathcal{O}^{\mathsf{\DSAL}}$ using $ \pp $ with audit log $\ledger$.  At each time step, $ \adv $ adaptively specifies a query $ q \in \{\mathsf{CreateAddress}$, $\Store$, $\AssignOwner$, $\Share$, $\Access\} $ to $ \chal $. $\chal$ forwards $ q $ to $\mathcal{O}^{\mathsf{\DSAL}}$, receives a response and forwards it to $ \adv $. $\chal$ also provides the view of $\set{L}$ to $ \adv $. $ \adv $ outputs a record $ \tx^* $ and let $\set{T}$ be the set of records with the same type of $ \tx^* $ that	$\mathcal{O}^{\mathsf{\DSAL}}$ generates during the query phase. 
		%
		
		If there exists a record $\tx \in \set{T}$ created in response to a query, $\adv$ wins and outputs 1 iff the following conditions hold -- (i) $m^* \ne m$;  (ii) $ \mathsf{VerifyLog}(\tx^*,\pp,\set{L}') = 1 $, where $ \set{L}' $ is the state of the audit log preceding $\tx$; iii) if $\tx^*$ is a record of ownership, $\sn^* = \sn$. Otherwise, it outputs 0.
%

$ \Pi $ is said to achieve \emph{record non-malleability} if $\Pr[{\trnm}_{\Pi,\adv}(\secparam) = 1] \leq \negl(\secparam)$.

\end{Definition}
\begin{Definition}[Audit Log Indistinguishability] \label{lind}
Let $ \Pi $ be a candidate \DSAL~scheme. Consider experiment  $\lind_{\Pi, \adv}(\secparam)$ between $ \adv $ and $\chal$. $\chal$ first samples a random bit $ b \sample \{0,1\}$, $\pp \gets \Setup(1^\secparam)$ and initializes two \DSAL~oracles $ \mathcal{O}^{\mathsf{\DSAL}}_0 $ and $ \mathcal{O}^{\mathsf{\DSAL}}_1 $, where each oracle $ \mathcal{O}^{\mathsf{\DSAL}}_i $ has a separate audit log $ \ledger_i $ as well as internal variables. At each time step, $ \adv $ adaptively specifies two queries $ Q,Q' $ of the same type (one of $\Store,\AssignOwner,\Share,\Access$). 
	$ \chal $ forwards $Q$ to $ \mathcal{O}^{\mathsf{\DSAL}}_0 $, $Q'$ to $ \mathcal{O}^{\mathsf{\DSAL}}_1 $ and receives the corresponding answers $R_0, R_1$. $ \chal $ forwards $ (R_b, R_{1-b}) $ to $ \adv $. $ \chal $ also provides to $ \adv $ two ledgers $ \ledger_{\mathsf{left}}= \ledger_b $, $ \ledger_{\mathsf{right}}= \ledger_{1-b} $, where $\ledger_b$ is the current audit log in $ \mathcal{O}^{\mathsf{\DSAL}}_b $. $ \adv $ outputs a bit $ b'  \in \{0, 1\}$. $ \adv $~wins if $ b' = b $ and outputs $ 1 $ else 0.

$ \Pi $ is said to achieve \emph{audit log indistinguishability} if
$\Pr[{\lind}_{\Pi,\adv}(\secparam)=1] \leq \frac{1}{2} + \negl(\secparam)$.

\end{Definition}

\label{ch:models}
   	


%



\section{Our Proposed Method}
\subsection{Overview}
$ \sys $ supports three data operations -- store, share, and access, wherein each operation is recorded in a distributed audit log (i.e., blockchain).
%
%
%
\begin{figure}[t]
    \centering
    \includegraphics[width=\columnwidth]{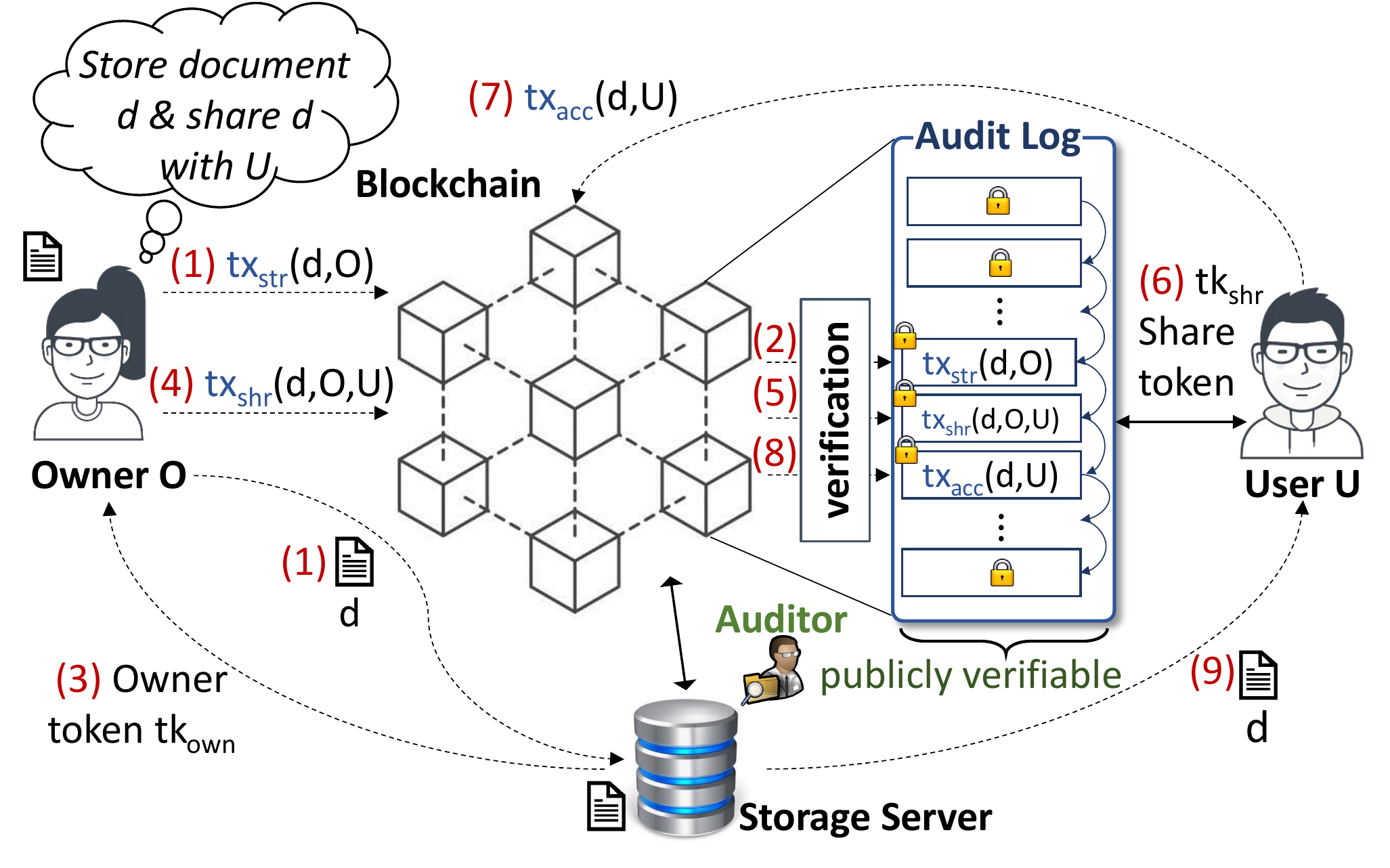}
    \caption{Overview of \sys~scheme.}
    \label{fig:overview} \vspace{-5mm}
\end{figure}
The high-level workflow (ref. \autoref{fig:overview})
is as follows, given a data $d$ to be stored on $\provider$,
$\owner$  creates ``store'' record \textcolor[HTML]{c82020}{(1)} and a token for  $\provider$ containing the message digest of $d$, while sending $d$ off-chain to $\provider$ for storage. 
The nodes in the blockchain then verify the validity of the store record \textcolor[HTML]{c82020}{(2)} before recording it to the audit log.
%
%
$\provider$ then creates an ``ownership'' record that certifies the ownership of $\owner$ on the data and provides a token \textcolor[HTML]{c82020}{(3)} that permits $\owner$ to share her data with $\user$ later.
To share the data, $\owner$ creates a ``share'' record \textcolor[HTML]{c82020}{(4)} demonstrating her ownership, which is then verified by the blockchain before being appended to the audit log \textcolor[HTML]{c82020}{(5)}.
The user then accesses this record to obtain a share token $\tokenshare$ \textcolor[HTML]{c82020}{(6)} indicating the authorization to access the shared data stored on $\provider$ for a limited period of time.
To access the shared data, the user creates an ``access'' record \textcolor[HTML]{c82020}{(7)} demonstrating the access is authorized by the owner, and an access token for $ \provider $ to serve the request. 
Finally, 
the record  \textcolor[HTML]{c82020}{(8)} is verified by the blockchain and included in the audit log.
If it is valid, $\provider$ sends data \textcolor[HTML]{c82020}{(9)} to the requested user $\user$.

To ensure all the data operations are confidential (e.g., hides participant identity or which data is being shared/accessed), while permitting public verification, we harness cryptographic commitment, encryption, and zero-knowledge techniques inspired by Zerocash \cite{zerocash_sp14}.

\subsection{Detailed Construction}

\subsubsection{Initialization} \label{sec.initialization}
The initialization generates public parameters for NP-statements to be proven/verified in zero-knowledge.
We initialize four empty Merkle trees with roots ($\rt_\store,\rt_\own,\rt_\share,\rt_\access$), each containing commitments of the corresponding data operation being performed.
We assume $ \sys $ permits up to $ N $ number of operations per type.

\begin{figure}
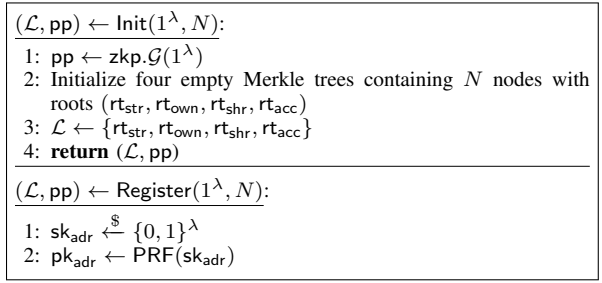

	\footnotesize
    \centering
	\fbox{\parbox{0.95\linewidth}{ 
		\underline{$ (\ledger, \pp) \gets \Init(1^\secparam, N)$}: 
		\begin{algorithmic}[1]
			\State $\pp \gets \zkKeyGen(1^\secparam)$ 
			\State Initialize four empty Merkle trees containing $ N $ nodes with roots $(\rt_\store, \rt_\own, \rt_\share, \rt_\access)$
			\State $\ledger \gets \{\rt_\store, \rt_\own, \rt_\share, \rt_\access\}$
			\State \Return ($\ledger, \pp$)
		\end{algorithmic}
		
    \algrule
	\footnotesize

		\underline{$ (\ledger, \pp) \gets \CreateAddress(1^\secparam, N)$}: 
		\begin{algorithmic}[1]
			\State $\addsk \sample \{0,1\}^\secparam$ 
			\State $\addpk \gets \PRF(\addsk)$
		\end{algorithmic}

	}}
	\caption{Initialization.}\label{alg.init}
\vspace{-5mm}
\end{figure}

\autoref{alg.init} presents the initialization in detail. 
Since \sys~operates on a distributed audit log, 
each participant needs an address keypair for communication and logging their data operations by using $\CreateAddress$.

Since $(\addpk,\addsk)$ is never revealed, 
each participant is free to re-use a single address for all operations. 

\subsubsection{Data Store Protocol} \label{sec.storage}
\autoref{alg.tx_store} presents our data store protocol.
Given data $ D $, $\owner$ first creates a commitment formed by the hash of the data $\CRH(D)$, $\provider$'s identity $ \addpk^\provider $ and a random value $ \rho $ as $ \cm_\store \gets  \Com_{r}(\CRH(D)||\addpk^\provider||\rho,\pp)$ (\autoref{alg.tx_store}, line \ref{fig:sys:DataStore:comGen}).
$\addpk^\provider$ is needed in $\cm_\store$ to tie store request of $ D $ with $\provider$.
On the other hand, the random $ \rho $ is used to ensure that the request can only be handled \emph{once} (see \autoref{sec.assignowner}).

$\owner$ generates a one-time unique signing key pair $(\sigpk,\sigsk) $ to sign their store request (\autoref{alg.tx_store}, line \ref{34}). This ensures unlinkability between multiple requests from $\owner$.
To achieve non-repudiation and prevent malleability attacks, such keys must be tied with $\addpk^\owner$.
Thus, $\owner$ creates a MAC tag $ h $ of the signing key $ \sigpk $ under their address private key $ \addsk^\owner $ as $h \gets \PRF(\addsk^\owner,h_\sig)$, where $h_\sig \gets \CRH(\sigpk)$ (\autoref{alg.tx_store}, line \ref{35}).
For creating a store record,
$\owner$ needs to show that the components of $(\cm_\store,h)$ have been formed correctly under their secrets $(\addpk^\provider, r, \rho, \filehash, \addsk^\owner, \h_\sig) $.
%

\PP{NP-statements for Data Store}
Given a store instance $ x = (\cm_\store,h)$, the witness $ w =(\addpk^\provider, r, \rho, \filehash, \addsk^\owner, h_\sig) $ is a valid witness for $ x $ if the following holds --
\begin{itemize}[leftmargin=.15in, topsep=1pt, itemsep = .3em]
	\item The commitment $ \cm_\store $ is well-formed, i.e., $\cm_\store = \Com_r(\CRH(D)||\addpk^\provider||\rho,\pp)$.
	
	\item The address secret key $\addsk^\owner$ ties $h_\sig$ to $h$, i.e., $h = \PRF(\addsk^\owner || h_\sig)$. 
	
\end{itemize}

$\owner$ then creates a store token $\tokenstore \gets \encryption.\Enc(\addpk^\provider,r||\rho||\filehash||\cm_\store)$ containing the openings of the commitment and encrypts it under $\provider$'s public key (\autoref{alg.tx_store}, line \ref{39}).
This token will be used by $\provider$ to respond to $\owner$'s request appropriately.
$\owner$ forms a store record $\tx_\store$, which contains the commitment $ \cm_\store $, the MAC tag $ h $, the encrypted token $ \tokenstore $, proof $\pi_\store$, and the signature verification key $ \sigpk $.
We store the token in the record to help with auditing. During an audit, $\owner$ simply uses their secret key to decrypt the token and verify record contents.
Finally, $\owner$ signs the record with $\sigsk$, and broadcasts it to the distributed audit log while sending the data $ D $ to $\provider$.
$\provider$ checks if the decrypted store token $ \tokenstore $ is consistent with the commitment and the record is valid (i.e., valid signature and valid proof).
If so, $\provider$ stores $D$ in an appropriate location and creates a record to give the ownership to $\owner$.
\begin{figure}[t]\centering
	\footnotesize

	\fbox{\parbox{0.95\linewidth}{ 
		\underline{$ (\tokenstore,\tx_{\store}) \gets \Store( \file, \addsk^\owner, \addpk^\provider,\pp$)}:
        \begin{algorithmic}[1]
				\State  $ \filehash \gets \CRH(\file) $ 

				\State  $\rho \sample \{0,1\}^\secparam$

				\State  $ \cm_\store \gets \Com_r(\addpk^\provider || \rho || \filehash,\pp)$ \label{fig:sys:DataStore:comGen}
				\State  $(\sigpk, \sigsk) \gets \signature.\keygen(1^\secparam)$ \label{34}

				\State  $\h_{\sig} \gets \CRH(\sigpk) $, \hs $\h \gets \PRF({\addsk^\owner}||\h_{\sig})$ \label{35}
				
				\State  $ x = (\cm_\store,h) $\label{36}
				\State $ w =  (\addpk^\provider, r, \rho, \filehash, \addsk^\owner, h_\sig) $\label{37}
				
				\State  $ \pi_\store \gets \zkProve(x,w,\pp) $\label{38}
				\State  $\tokenstore \gets \encryption.\Enc(\addpk^\provider,r||\rho||\filehash||\cm_\store)$\label{39}

				\State $m = x||h_\sig||\pi_\store||\tokenstore$\label{310}
				
				\State  $\sigma \gets \signature.\Sig(\sigsk, m)$ \label{311}
				\State $ \tx_{\store} = (\sigpk, \sigma, \cm_{\store}, h,  \pi_\store, \tokenstore)$ \label{312}
				\State \Return $(\tokenstore,\tx_{\store})$ \label{313}
	\end{algorithmic}
    \algrule
    \FuncHeader{$(\tokenowner,\tx_{\assign}) \gets \AssignOwner(\tokenstore, \addsk^\provider, \addpk^\owner,\pp$)}:
		\begin{algorithmic}[1]
		        \setcounter{ALG@line}{13}
				\State  $\sn \gets \PRF({\addsk^\provider}||\rho)$ \label{fig:sys:assignowner:genSN}
				\State  $ \cm_\own\gets \Com_{r'}(\addpk^\owner || \filehash',\pp)$ \Comment{$ \filehash'=\filehash $}\label{315}

				\State  $ \mypath_\store = $ Merkle path from $\cm_{\store}$ to $ \rt_\store $ ( $\rt_\store$ from $\ledger$)\label{316}
				\State  $(\sigpk, \sigsk) \gets \signature.\keygen(1^\secparam)$ \label{317}
								
				\State  $\h_\sig \gets \CRH(\sigpk)$, \hs $\h \gets \PRF({\addsk^\provider}||\h_{\sig})$ \label{318}
				
				\State  $ x = (\rt_\store, \sn, \cm_\own,h) $\label{319}
				\State $ 
				w =(\mypath_\store, \addsk^\provider, \rho, r, \filehash, \addpk^\provider, r', \filehash', \addpk^\owner, h_\sig) $\label{320}
				
				\State  $ \pi_\own \gets \zkProve(x,w,\pp) $\label{321}
				\State  $\tokenowner \gets \encryption.\Enc(\addpk^\owner,r'||\filehash'||\cm_{\own})$ \label{322}
				\State $m = x||h_\sig||\pi_\assign||\tokenowner$\label{323}
				
				\State  $\sigma \gets \signature.\Sig(\sigsk, m)$ \label{324}
				\State $ \tx_{\assign} = (\rt_\store, \sigpk, \sigma,  \sn, \cm_\own, \h, \pi_\own, \tokenowner )$ \label{325}
				\State \Return $(\tokenowner,\tx_{\assign})$\label{326}
		\end{algorithmic}

}}
\caption{Data store and owner assignment protocols.}\label{alg.tx_store}\label{alg.tx_assign}
\vspace{-5mm}
\end{figure}

\subsubsection{Assign Ownership Protocol} \label{sec.assignowner}

$\provider$ performs an assign ownership operation on receiving a $\tokenstore$. 
$\provider$ decrypts $\tokenstore$ to identify which data needs to be assigned. 
A serial number $\sn$ is generated using the seed $\rho$ (\autoref{alg.tx_assign}, line \ref{fig:sys:assignowner:genSN}) in the store record.
$\sn$ uses a collision-resistant $\PRF$ function to ensure that only one assignment is created for the corresponding store operation.
Without this, $\provider$ can use a single $\tokenstore$ to maliciously perform assign operations to multiple data owners.
Subsequently, $\provider$ generates a commitment $\cm_\assign \gets \Com_{r'}(\pk_\addr^\owner || v',\pp)$ (\autoref{alg.tx_store}, line \ref{315}) to represent ownership assignment. 
To successfully create an assign record, $\provider$ generates a proof of ownership $\pi_\own$ (\autoref{alg.tx_store}, line \ref{321}). 
Verification of the proof shows that $\provider$ received a store request for the corresponding data. 
The proof is generated by the following NP statements --

\PP{NP-statements for Data Owner Assignment}
Given an assign owner instance \sloppy $ x = (\rt_\store, \sn, \cm_\own,h)  $, the witness $ w = (\mypath_\store, \addsk^\provider, \rho, r, \filehash, \addpk^\provider, r', \filehash', \addpk^\owner, \h_\sig)  $ is a valid witness for $ x $ if the following holds --
\begin{itemize} [leftmargin=.15in, topsep=1pt, itemsep = .3em]
	\item Commitment $ \cm_\store $  appears in the Merkle tree $ \rt_\store $, i.e., $\rt_\store = \CRH(\dots(\CRH(\CRH(\cm_\store) || \mypath_\store^{d}) || \mypath_\store^{d-1}) || \dots || \mypath_\store^{1}))$  where $d$ is the height of the Merkle tree and $\mypath_\store^i$ represents the sibling node at height $i$.
	
	\item Commitments $ \cm_\store $ and $ \cm_{\own} $ are well-formed.
	
	\item Serial number $ \sn $ is computed correctly, i.e., $\sn = \PRF(\addsk^\provider||\rho)$  
	\item Values committed to $ \cm_\store $ and $ \cm_{\own}$ are the same.
	
	\item The address secret key $\addsk^\provider$ ties $h_\sig$ to $h$.
	
\end{itemize}

We utilize Merkle trees to prove set membership i.e, the statement to prove $\cm_\store$ appears in $\rt_\store$. 
The tree is stored locally by the blockchain nodes and only the root is added to the record. 
This is because we want to capture the state of the tree when the particular record was created. 
$\provider$ then creates an encrypted ownership token $\tokenowner$ (\autoref{alg.tx_store}, line \ref{322}).
Finally, $\provider$ signs record $\tx_\assign$, using a one-time signing key, and appends it to the distributed audit log.
$\provider$ also sends $\tokenowner$ to $\owner$.

		

\subsubsection{Sharing Protocol}\label{sec.sharing}

Given an ownership token $\tokenowner$, a share expiry timestamp $\ts$, $\user$'s public identity $\addpk^\user$ and $\owner$'s identity $\addsk^\owner$ a share operation generates a share token $\tokenshare$ to $\user$ and a record $\tx_\share$. 
A commitment $\cm_\share$ is created to $\user$, share expiry timestamp and the hash of the data being shared (\autoref{alg.tx_share}, line \ref{fig:sys:Share:genCm}).
$\owner$ generates a valid proof $\pi_\share$ (\autoref{alg.tx_share}, line \ref{47}) by proving the following NP statements --

\PP{NP-statements for Data Sharing}
Given a share instance $ x = (\rt_\own, \cm_{\share},h)  $, the witness $ w = (\mypath_\own, \addpk^\owner, r, \filehash, \addpk^\user, r', \filehash', \h_\sig, \addsk^\owner) $ is a valid witness for $ x $ if the following holds --
\begin{itemize}	[leftmargin=.15in, topsep=1pt, itemsep = .3em]
	\item Commitment $ \cm_\own $  appears in the Merkle tree $ \rt_\own $.
	
	\item Commitments $ \cm_\own $ and $ \cm_{\share} $ are well-formed.  
	
	\item Values committed to $ \cm_\store $ and $ \cm_{\own} $  are the same.  
	
	\item The expiration timestamp committed to $ \cm_\share $ is valid, i.e., $\ts \ge 0$.
	 
	\item The address secret key $\addsk^\owner$ ties $h_\sig$ to $h$.

\end{itemize}
Verification of this proof means that $\owner$ owns the corresponding data being shared and is sharing it for a valid period of time. 
Once the proof is generated, all the generated data is bundled into a record $\tx_\share$ and signed using $\sigsk$ (\autoref{alg.tx_share}, line \ref{410}).
The record is added to the distributed audit log for verification and the token $\tokenshare$ is sent to $\user$.
Note that $\owner$ can share their data multiple times because they can re-use their token $\tokenowner$ more than once.

\begin{figure}[t]
	\centering
	\fbox{\parbox{0.95\linewidth}{
	\footnotesize
		\FuncHeader{$ (\tokenshare, \tx_{\share} )\gets \Share(\tokenowner, \addsk^\owner, \addpk^\user, \ts,\pp)$}:
		\begin{algorithmic}[1]
				\State  $ \cm_\share \gets \Com_{r'}(\addpk^\user || \ts || \filehash',\pp)$ \label{fig:sys:Share:genCm} \Comment{$ v' = v $}

				\State  $\mypath_\own \gets $ path from $\cm_\own$ to $\rt_\own$ ( $\rt_\own$ from $\ledger$) \label{42}
				\State  $(\sigpk, \sigsk) \gets \signature.\keygen(1^\secparam) $	\label{43}
				\State  $\h_{\sig} \gets \CRH(\sigpk)$, \hs $\h \gets \PRF({\addsk^\owner}||\h_{\sig})$\label{44}
				\State  $ x = (\rt_\own, \cm_{\share},h)  $\label{45}
				\State $ w = (\mypath_\own, \addpk^\owner, r, \filehash, \addpk^\user, r', \filehash', h_\sig, \addsk^\owner) $ \label{46}
				\State  $\pi_{\share} \gets \zkProve(x,w,\pp)$ \label{47}

				\State  $\tokenshare \gets \encryption.\Enc(\addpk^\user,r'||\filehash'||\ts|| \cm_\share)$ \label{48}
				\State $m = x||h_\sig||\pi_\share||\tokenshare$\label{49}
				\State  $\sigma \gets \signature.\Sig(\sigsk, m)$ \label{410}
				\State $ \tx_{\share} = (\rt_\own, \sigpk,\sigma, \cm_{\share},  \h, \pi_\share, \tokenshare) $ \label{411}
				\State \Return $(\tokenshare,\tx_{\share})$\label{412}
        \end{algorithmic}
        \algrule
        	\FuncHeader{$ (\tokenaccess,\tx_\access) \gets \Access(\tokenshare,\addsk^{\user},\addpk^{\provider},\pp)$}:
        \begin{algorithmic}[1]
                \setcounter{ALG@line}{12}
				\State $ \cm_\access\gets \Com_{r'}(\addpk^\provider || \ts-\curr || \filehash',\pp)$ \label{fig:sys:Access:genCm} \Comment{$ v'=v $}
	
				\State  $\mypath_\share$ is the path from $\cm_\share$ to $\rt_\share$ in Merkle tree ($\rt_\share$ from $\ledger$)	\label{414}
			
				\State $(\sigpk, \sigsk) \gets \signature.\keygen(1^\secparam)$\label{415}
				\State $\h_{\sig} \gets \CRH(\sigpk)$, \hs $\h \gets \PRF(\addsk^{\user}||\h_{\sig})$  \label{416}
				\State $ x = (\rt_\share, \cm_{\access}, h)  $\label{417}
				\State $ w = (\mypath_\share, \addpk^\user, r, \filehash, \addpk^\provider, r', \filehash', \ts, \curr, h_\sig, \addsk^\user) $ \label{418}

				\State $\pi_{\access} \gets \zkProve(x,w,\pp)$ \label{419}
				
				\State $\tokenaccess \gets \encryption.\Enc(\addpk^\provider, r'||\filehash'|| \curr|| \cm_\access)$\label{420}

				\State $m = x||h_\sig||\pi_\access||\tokenaccess$\label{421}
				
				\State  $\sigma \gets \signature.\Sig(\sigsk, m)$ \label{422}
				
				\State $\tx_{\access} = (\rt_\share, \sigpk, \sigma, \cm_\access, \cm_\share, \ts, \h, \pi_{\access}, \tokenaccess)$ \label{423}
				
				\State \Return $(\tokenaccess,\tx_\access)$\label{424}
        \end{algorithmic}
}}
	\caption{Data share and access protocols.}\label{alg.tx_share} \label{alg.tx_access}
\vspace{-7mm}
\end{figure}

\subsubsection{Access Protocol} \label{sec.access}
$\user$ uses $\tokenshare$ obtained from $\owner$ to perform an access operation.
A commitment representing an access $\cm_\access \gets \Com_{r'}(\addpk^\provider||\fid',\pp)$ is generated using $\provider$'s public key and the hash of the data that needs to be accessed (\autoref{alg.tx_access}, line \ref{fig:sys:Access:genCm}).
$\user$ then creates a proof $\pi_\access$ by proving the following -- 

\PP{NP-statements for Data Access}
Given an access instance $ x = (\rt_\share, \cm_{\access},h)  $, the witness $ w = (\mypath_\share, \addpk^\user, r, \filehash, \addpk^\provider, r', \filehash', \ts, 
\curr, \h_\sig, \addsk^\user)  $ is a valid witness for $ x $ if the following holds --
\begin{itemize}	[leftmargin=.15in, topsep=1pt, itemsep = .3em]
	\item Commitment $ \cm_\share $  appears in Merkle tree $ \rt_\share $.
	
	\item Commitments $ \cm_\share $ and $ \cm_{\access} $ are well-formed.  
	
	\item Values committed to $ \cm_\share $ and $ \cm_{\access} $  are the same.
	
	\item The timestamp committed to $ \cm_\access $ is not expired, i.e., $0 < \curr < \ts$ where $\curr$ is the current timestamp.
 
	\item The address secret key $\addsk^\user$ ties $h_\sig$ to $h$.
\end{itemize}
A valid proof for the above statements can only be generated if $\user$ was given the share token by $\owner$ and the share expiry timestamp is not expired.
We note that $\user$ can access the data as many times as they wish within the given time period after which they will not be able to generate a valid $\pi_\access$.
Finally, $\user$ creates an access record (\autoref{alg.tx_share}, line \ref{423}) and appends it to the distributed audit log.
$\user$ also sends the access token to the $\provider$ to obtain the data shared with them. 
$\provider$ sends the data only if the record (i.e $\tx_\access$) corresponding to $\tokenaccess$ has been validated by the distributed audit log. 
\subsubsection{Record Verification}
The $\VerifyLog$ protocol (see \autoref{alg.tx_verify}) runs on the distributed audit log nodes to verify the zero-knowledge proof and the signature included in the record.
If either of the verification procedures fails, the record is discarded.
If the record verification succeeds, the audit log nodes update their local copies of the corresponding commitment tree and appends the record to the audit log. 
%

	

\begin{figure}[t]
	\centering
	\fbox{\parbox{0.95\linewidth}{ 
	\footnotesize
		\FuncHeader{$ \{0,1\} \gets \VerifyLog(\tx,\ledger,\pp)  $}: 
        \begin{algorithmic}[1]
            \item \textbf{if} $ \tx = (\sigpk, \sigma, \cm_{\store}, h,  \pi_\store, \tokenstore) $ \Comment{$\tx_\store$}
            \item \hs $ \x = (\cm_\store, \h)$ 
            \item \hs $ m = x||h_\sig||\pi_\store||\tokenstore$
            \item \textbf{else if} $ \tx = (\rt_\own, \sigpk, \sigma,  \sn, \cm_\store, \h, \pi_\own, \tokenowner) $ \Comment{$ \tx_\assign$}
            \item \hs If $\sn \in \ledger$, abort.
            \item \hs $ \x = (\rt_\store, \sn, \cm_\own,h)$,
            \item \hs $ m = x||h_\sig||\pi_\assign||\tokenowner$
            
            \item \textbf{else if} $ \tx= (\rt_\share, \sigpk,\sigma, \cm_{\share},  \h, \pi_\share, \tokenshare) $ \Comment{$\tx_\share$}
            \item \hs $ \x = (\rt_\own, \cm_\share, h)$
            \item \hs $ m = x||h_\sig||\pi_\share||\tokenshare$
           	\item \textbf{else if} $ \tx = (\rt_\access, \sigpk, \sigma, \cm_\access, \h, \pi_{\access}, \tokenaccess) $ \Comment{$\tx_\access$}
            \item \hs $ \x = (\rt_\share, \cm_\access,h)$
            \item \hs  $  m = x||h_\sig||\pi_\access||\tokenaccess$
            \item \textbf{return} $\zkVerify(\x, \pi,\pp) \land \signature.\Verify({\sigpk}, m,\sigma)$

		\end{algorithmic}
		}}
\caption{Verification protocol.}\label{alg.tx_verify}
\vspace{-7mm}
	
\end{figure}
 \label{ch:method}
   	\section{Security Analysis}
\begin{Theorem}
    \sys~satisfies record non-malleability by \autoref{lnm}.
\end{Theorem}
\begin{proof}[Proof (sketch)]  Let $Q_\CreateAddress = \{1^\lambda, 1^\lambda, \dotsc\}$ be the queries by $\adv$ to $\CreateAddress$ and $R_\CreateAddress = \{(\addpk^{\provider,1}, \addsk^{\provider,1}), (\addpk^{\provider,2}, \addsk^{\provider,2}), \dotsc\}$ be $\chal$'s response. 

Without loss of generality, let
    \sloppy $Q_\AssignOwner=\{(\tokenstore^1,\addsk^{\provider,1},\addpk^{\owner,1},\pp), (\tokenstore^2,\addsk^{\provider,2},\addpk^{\owner,2},\pp),, \dotsc\}$ be the queries created by $\adv$ and $R_\AssignOwner=\{(\tokenowner^1, \tx_\own^1), (\tokenowner^2, \tx_\own^2), \dotsc\}$ be the response by $\chal$. $\adv$ wins the experiment $\trnm$ when it outputs a record $ \tx_\own^* $ such that (i) $m^* \neq m$, where $m$ is the message in some $\tx_\own \in R_\AssignOwner$;  (ii) $ \mathsf{VerifyLog}(\tx_\own^*,\set{L}',\pp) = 1 $, where $ \set{L}' $ is the state of the audit log preceding $\tx_\own$; (iii) $\sn^*=\sn$ iff $\tx^*$ is an ownership record.

  Assume that  $\adv$ wins by creating a record $\tx_\own^*=(\rt_\store^*, \sigpk^*, \sigma^*,  \sn^*, \cm_\own^*, \h^*, \pi_\own^*, \tokenowner^*)$ with proof instance $ x^* = (\rt_\store^*, \sn^*, \cm_\own^*, h^*)$ and $m^* = x^*||h_\sig^*||\pi_\own^*||\tokenowner^*$. We consider the following events.

    $E_1:$ $\adv$ wins, and there exists $\pk_\sig \in R_\AssignOwner$ such that $\pk_\sig^* = \pk_\sig$. To win, $\adv$ produces $\sigma^*$ such that $\Sigma.\Verify(\pk_\sig^*, m^*, \sigma^*) = 1$, $m^* \ne m$ and $\pk_\sig^* = \pk_\sig$. However, due to the SUF-CMA property of the signature scheme \cite{schnorr}, this happens with negligible probability.  
    
    $E_2:$ $\adv$ wins, $E_1$ does not occur, and there exists $\pk_\sig \in R_\AssignOwner$ such that $h^*_\sig = h_\sig$. When $\adv$ wins, it crafts a $h^*_\sig = h_\sig$ when $\pk_\sig^* \ne \pk_\sig$. However, the hash function \cite{mimc} is collision-resistant, a contradiction. 
    
    $E_3:$ $\adv$ wins, $(E_1, E_2)$ do not occur and  $h^* = \PRF(\addsk^{\provider^*}|| h^*_\sig)$ for some $\addsk^{\provider^*} \in R_\CreateAddress$. Due to the one-wayness property of $\PRF$ (used to generate $\addpk^{\provider^*}$), $\adv$ will not be able to deduce $\addsk^{\provider^*}$ and generates $h^*$ with a random key. However, due to indistinguishability property of $\PRF$ (used to generate $h^*$), $\adv$ wins negligbly.
    
    $E_4:$ $\adv$ wins, $(E_1, E_2, E_3)$ do not occur, $\tx^*$ is  an ownership record and  $h^* \ne \PRF(\addsk^{\provider^*}|| h^*_\sig)$ for \textit{all} $\addsk^{\provider^*} \in R_\CreateAddress$. If $\adv$ wins, it outputs a modified record such that $\sn^*=\sn$. However, since in $E_4$ $\addsk^{\provider^*} \ne \addsk^{\provider}$, it means that to generate $\sn^*$, $\adv$ found a collision in the $\PRF$ needed to generate $\sn$ which occurs with negligible probability due to the collision-resistant property of the $\PRF$ function used to generate $\sn$. 
    \qedhere
\end{proof}

\begin{Theorem}
	 \sys~satisfies audit log indistinguishability by \autoref{lind}
\end{Theorem}
\begin{proof}[Proof (sketch)]
	$\chal$ creates a simulation $\lind_\Sim$ that can answer the queries ($Q, Q'$) independent to the selected bit $ b $ as follows. $\chal$ modifies the commitment $\cm_\store^* \gets \Com_{r}(k,\pp)$ by using random inputs $k, r$. To answer $\AssignOwner$ queries, $\chal$ modifies $\AssignOwner$ to create $\sn$ and $h$ randomly. $\cm_\own$ is computed like in $\Store$. $\tokenowner^* \gets \encryption.\Enc(e,p)$ is generated using a random plaintext $p$ and key $e$. Proof $\pi_\own^* \gets \zkSim(x,\pp)$ is computed using a simulator. To answer $\Share$ or $\Access$ queries, $\chal$ computes $h$, commitments ($\cm_\share$/$\cm_\access$), tokens ($\tokenshare$/$\tokenaccess$), proofs ($\pi_\share$/$\pi_\access$) by following modifications in $\AssignOwner$. 

We prove this by constructing a sequence of hybrid games and show that they are indistinguishable as follows. 
Let $G_0$ be the original protocol $\lind_\Pi$. We define $G_1$ as similar to $G_0$, except that $\chal$ modifies the commitment in the response to use a $\cm^* \gets  \Com_{r}(k, \pp)$ by using random inputs $k, r$. We have that $ G_1 \stackrel{c}{\approx} G_0 $ due to  the hiding property of the commitment scheme.%
We define $G_2$ as similar  to $G_1$, except that the token in the response is replaced with $\token^* \gets \encryption.\Enc(e, p)$ where $e$ is a random public key and $p$ is a random plaintext.
We argue that $ G_2 \stackrel{c}{\approx} G_1 $ due to the IND-CCA and IK-CCA properties of the encryption scheme \cite{dhaes}.
We define $G_3$ as similar to $G_2$, except that all $\PRF$ generated values (i.e, $h, \sn$) in the response are replaced with values generated using a random function.
We say that $ G_3 \stackrel{c}{\approx} G_2 $ due to the  indistinguishability between $\PRF$ and random number function.
Finally, we define $ G_4 $ as similar to $G_3$, except that the zero-knowledge proof in the response is replaced by a simulated proof $\pi^* \gets \zkSim(x, \pp)$.
We have that $ G_4 \stackrel{c}{\approx} G_3 $ due to the zero-knowledge property of the proof system.

We can see that  $G_4$ is the simulation  $\lind_\Sim$. 
Therefore, $\lind_\Pi = G_0 \stackrel{c}{\approx} G_1 \stackrel{c}{\approx} G_2 \stackrel{c}{\approx} G_3 \stackrel{c}{\approx} G_4 = \lind_\Sim$.
Since $\lind_\Sim$ can simulate the response to $Q$ and $Q'$ independent of bit $b$, $\adv$ can win $\lind_\Pi$ game with negligible probability.
\qedhere
\end{proof} \label{ch:sec_anal}
\section{Implementation} \label{sec.implementation}
We implemented $\sys$~in Rust with 3,100 lines of code.
We used Hyperledger Fabric \cite{fabric} to set up the audit log as the distributed ledger.
Each data operation was implemented as Hyperledger chaincode (equivalent to smart contracts) in Go. 
For the signature scheme, we implemented Schnorr signature scheme \cite{schnorr}. 
For the asymmetric encryption, we implemented ECIES with AES-GCM as the underlying symmetric encryption scheme \cite{dhaes}. 
For the commitment scheme, we used Pedersen commitment \cite{pedersen}.
We implemented the Merkle trees with Poseidon hash function as it can efficiently compute hashes of two elements at once \cite{poseidon}.
%
We implemented MiMC as the primary hash function as well as PRF due to its low multiplicative complexity \cite{mimc}.
Finally, our implementation made use of Bulletproofs \cite{bulletproof} as the back-end zero-knowledge proof due to its small proof size and transparent setup. 
We prove/verify statements in zero-knowledge for each data operation as follows --

\PP{Store} We implemented a circuit with one Pedersen commitment opening gadget $\PedG$ (based on \cite{boneh}) to show knowledge of openings of a commitment and one hash pre-image gadget $\HshG$ (based on \cite{mimc}) to show knowledge of a pre-image of a MiMC hash. 


\PP{Assign} We implemented a circuit with one $\MemG$, two $\PedG$, and two $\HshG$. Here, $\MemG$ is a Merkle tree based proof of membership gadget.

\PP{Share} We implemented a circuit with one range gadget $\RgeG$, one $\HshG$, one $\MemG$, and two $\PedG$ gadgets. $\RgeG$ (based on \cite{bulletproof}) is used to show knowledge that a value ($\ts$ in our case) is within a given range. 

\PP{Access} We 
implemented a circuit with $\RgeG$, one $\HshG$, one $\MemG$, and two $\PedG$ gadgets.

\label{ch:impl}
   	
\section{Experimental Evaluation} \label{sec.experiments}

\begin{figure*}[!ht]
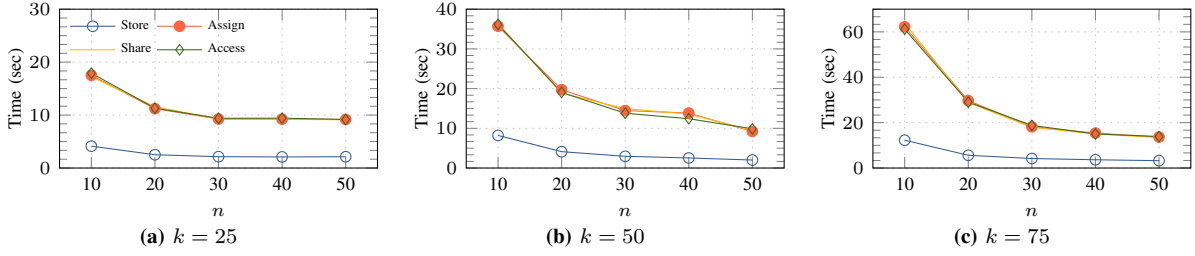
  
    \begin{subfigure}[b]{0.32\textwidth}
        \centering
        \input{fig/exp/tikz/latency_tx_fixed_25.tex}
         \vspace{-2mm}
        \caption{$k=25$}
        \label{fig:latency_tx_fixed_25}
    \end{subfigure}
    \begin{subfigure}[b]{0.32\textwidth}
        \centering
        \input{fig/exp/tikz/latency_tx_fixed_50.tex}
         \vspace{-2mm}
        \caption{$k=50$}
        \label{fig:latency_tx_fixed_50}
    \end{subfigure}
    \begin{subfigure}[b]{0.32\textwidth}
        \centering
        \input{fig/exp/tikz/latency_tx_fixed_75.tex}
         \vspace{-2mm}
        \caption{$k=75$}
        \label{fig:latency_tx_fixed_75}
    \end{subfigure}%
\caption{Verification latency.}\label{fig:exp:verificationlatency} \vspace{-5mm}
\end{figure*}
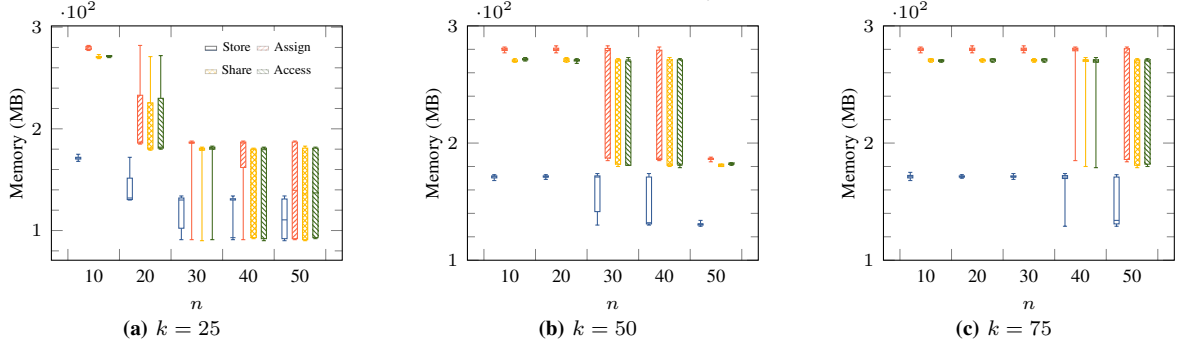
\begin{figure*}[!ht]  

    \begin{subfigure}{0.33\textwidth}
      \centering
        \pgfmathdeclarefunction{fpumod}{2}{%
    \pgfmathfloatdivide{#1}{#2}%
    \pgfmathfloatint{\pgfmathresult}%
    \pgfmathfloatmultiply{\pgfmathresult}{#2}%
    \pgfmathfloatsubtract{#1}{\pgfmathresult}%
    \pgfmathfloatifapproxequalrel{\pgfmathresult}{#2}{\def\pgfmathresult{5}}{}%
}
\begin{tikzpicture}
  \begin{axis}
    [width=\textwidth,
      cycle list={{store_color},{assign_color, pattern=custom north east lines, pattern color=assign_color},{share_color, pattern=crosshatchh, pattern color=share_color},{access_color, pattern=custom north west lines, pattern color=access_color}},
  legend pos = north west,
  legend entries = {Store, Assign, Share, Access},
  legend columns = 2,
  legend style={draw=none,font=\tiny,at={(0.5,0.98)},anchor=north west},
      boxplot/box extend=0.10,
        boxplot={
      %
      %
      draw position={1/5 + floor(\plotnumofactualtype/4) + 1/5*fpumod(\plotnumofactualtype,4)},
  },
  xtick={0,1,2,...,60},
  x tick label as interval,
  xticklabels={%
      {10},%
      {20},%
      {30},%
      {40},%
      {50},%
  },
  x tick label style={
      align=center
  },
    boxplot/draw direction=y,
    ylabel={Memory (MB)},
    ylabel style={font=\scriptsize},
    xlabel={$n$},
    minor y tick num=4,
        ylabel shift={-.5em},
    yticklabel style={font=\scriptsize},
    xticklabel style={font=\scriptsize},
    xlabel style={font=\scriptsize},
    ylabel style={font=\scriptsize},
    scaled y ticks = base 10:-2,
    legend image post style={scale=0.25, line width=0.0pt},
    ]
    \addlegendimage{area legend,draw opacity=0.5,pattern=horizontal lines,pattern color=store_color}
    \addlegendimage{area legend,draw opacity=0.2,pattern color=assign_color,pattern=custom north east lines}
    \addlegendimage{area legend,draw opacity=0.2,pattern=crosshatchh,pattern color=share_color}
    \addlegendimage{area legend,draw opacity=0.2,pattern=custom north west lines,pattern color=access_color}
    \addplot+[
    boxplot prepared={
      median=171.25,
      lower whisker=168,
      lower quartile=170.25,
      upper quartile=172,
      upper whisker=175
    },
    ]
    coordinates {};
    \addplot+[
    boxplot prepared={
      median=279.5,
      lower whisker=277,
      lower quartile=278,
      upper quartile=281,
      upper whisker=282
    },
    ] coordinates {};
    \addplot+[
    boxplot prepared={
      median=271,
      lower whisker=269,
      lower quartile=270,
      upper quartile=271,
      upper whisker=273
    },
    ] coordinates {};
    \addplot+[
    boxplot prepared={
      median=271,
      lower whisker=270,
      lower quartile=271,
      upper quartile=272,
      upper whisker=272
    },
    ] coordinates {};
    \addplot+[
    boxplot prepared={
      median=132,
      lower whisker=130,
      lower quartile=130.5,
      upper quartile=151.5,
      upper whisker=172
    },
    ] coordinates {};
    \addplot+[
    boxplot prepared={
      median=187,
      lower whisker=185,
      lower quartile=186,
      upper quartile=233,
      upper whisker=282
    },
    ] coordinates {};
    \addplot+[
    boxplot prepared={
      median=181,
      lower whisker=179,
      lower quartile=180,
      upper quartile=225.5,
      upper whisker=271
    },
    ] coordinates {};
    \addplot+[
    boxplot prepared={
      median=182,
      lower whisker=180,
      lower quartile=181,
      upper quartile=230,
      upper whisker=272
    },
    ] coordinates {};
    \addplot+[
    boxplot prepared={
      median=130,
      lower whisker=91,
      lower quartile=102.25,
      upper quartile=132,
      upper whisker=134
    },
    ] coordinates {};      
    \addplot+[
    boxplot prepared={
      median=186,
      lower whisker=91,
      lower quartile=186,
      upper quartile=187,
      upper whisker=188
    },
    ] coordinates {};
    \addplot+[
    boxplot prepared={
      median=180,
      lower whisker=90,
      lower quartile=179,
      upper quartile=181,
      upper whisker=182
    },
    ] coordinates {};
    \addplot+[
    boxplot prepared={
      median=181,
      lower whisker=91,
      lower quartile=180,
      upper quartile=182,
      upper whisker=183
    },
    ] coordinates {};

    \addplot+[
    boxplot prepared={
      median=93,
      lower whisker=91,
      lower quartile=130,
      upper quartile=131.25,
      upper whisker=134
    },
    ] coordinates {};
    \addplot+[
    boxplot prepared={
      median=186,
      lower whisker=91,
      lower quartile=162,
      upper quartile=187,
      upper whisker=188
    },
    ] coordinates {};
    \addplot+[
    boxplot prepared={
      median=180,
      lower whisker=92,
      lower quartile=93,
      upper quartile=180,
      upper whisker=181
    },
    ] coordinates {};
    \addplot+[
    boxplot prepared={
      median=179.5,
      lower whisker=90,
      lower quartile=92,
      upper quartile=181,
      upper whisker=182
    },
    ] coordinates {};

    \addplot+[
    boxplot prepared={
      median=110.5,
      lower whisker=90,
      lower quartile=92,
      upper quartile=131,
      upper whisker=134
    },
    ] coordinates {};      
    \addplot+[
    boxplot prepared={
      median=139.5,
      lower whisker=91,
      lower quartile=92,
      upper quartile=187,
      upper whisker=188
    },
    ] coordinates {};
    \addplot+[
    boxplot prepared={
      median=136,
      lower whisker=90,
      lower quartile=91,
      upper quartile=181,
      upper whisker=183
    },
    ] coordinates {};
    \addplot+[
    boxplot prepared={
      median=137,
      lower whisker=92,
      lower quartile=93,
      upper quartile=181,
      upper whisker=182
    },
    ] coordinates {};
  \end{axis}
\end{tikzpicture}
         \vspace{-2mm}
        \caption{$k=25$}
        \label{fig:memory25}
    \end{subfigure}
    \begin{subfigure}{0.33\textwidth}
      \centering
        \usetikzlibrary{patterns}


\pgfmathdeclarefunction{fpumod}{2}{%
    \pgfmathfloatdivide{#1}{#2}%
    \pgfmathfloatint{\pgfmathresult}%
    \pgfmathfloatmultiply{\pgfmathresult}{#2}%
    \pgfmathfloatsubtract{#1}{\pgfmathresult}%
    \pgfmathfloatifapproxequalrel{\pgfmathresult}{#2}{\def\pgfmathresult{5}}{}%
}
\begin{tikzpicture}
    \begin{axis}
    [width=\textwidth,
      cycle list={{store_color},{assign_color, pattern=custom north east lines, pattern color=assign_color},{share_color, pattern=crosshatchh, pattern color=share_color},{access_color, pattern=custom north west lines, pattern color=access_color}},
  legend pos = north west,
  legend style={legend cell align=left, align=left, fill=none, draw=none,inner sep=-0pt, row sep=0pt,at={(0,1.2)}},
  legend style={/tikz/every even column/.append style={column sep=1cm}},
      boxplot/box extend=0.10,
        boxplot={
      %
      %
      draw position={1/5 + floor(\plotnumofactualtype/4) + 1/5*fpumod(\plotnumofactualtype,4)},
  },
  xtick={0,1,2,...,60},
  x tick label as interval,
  xticklabels={%
      {10},%
      {20},%
      {30},%
      {40},%
      {50},%
  },
  x tick label style={
      align=center
  },
    boxplot/draw direction=y,
    ylabel={Memory (MB)},
    ylabel style={font=\scriptsize},
    xlabel={$n$},
    minor y tick num=4,
    ylabel shift={-.5em},
    yticklabel style={font=\scriptsize},
    xticklabel style={font=\scriptsize},
    xlabel style={font=\scriptsize},
    ylabel style={font=\scriptsize},
    scaled y ticks = base 10:-2,
    ymax = 300,
    ymin = 100,
    ytick distance=100,
    ]
      \addplot+[
      boxplot prepared={
        median=171.25,
        lower whisker=168,
        lower quartile=170.25,
        upper quartile=172,
        upper whisker=173
      },
      ]
      coordinates {};
      \addplot+[
      boxplot prepared={
        median=280,
        lower whisker=277,
        lower quartile=279,
        upper quartile=281,
        upper whisker=282
      },
      ] coordinates {};
      \addplot+[
      boxplot prepared={
        median=270,
        lower whisker=269,
        lower quartile=269.5,
        upper quartile=271,
        upper whisker=272
      },
      ] coordinates {};
      \addplot+[
      boxplot prepared={
        median=271,
        lower whisker=270,
        lower quartile=271,
        upper quartile=272,
        upper whisker=273
      },
      ] coordinates {};
      \addplot+[
      boxplot prepared={
        median=171,
        lower whisker=169,
        lower quartile=171,
        upper quartile=172,
        upper whisker=173
      },
      ] coordinates {};
      \addplot+[
      boxplot prepared={
        median=280,
        lower whisker=277,
        lower quartile=279,
        upper quartile=281,
        upper whisker=283
      },
      ] coordinates {};
      \addplot+[
      boxplot prepared={
        median=271,
        lower whisker=269,
        lower quartile=270,
        upper quartile=272,
        upper whisker=273
      },
      ] coordinates {};
      \addplot+[
      boxplot prepared={
        median=271,
        lower whisker=268,
        lower quartile=270,
        upper quartile=271,
        upper whisker=272
      },
      ] coordinates {};
      \addplot+[
      boxplot prepared={
        median=171,
        lower whisker=130,
        lower quartile=141.5,
        upper quartile=172,
        upper whisker=174
      },
      ] coordinates {};      
      \addplot+[
      boxplot prepared={
        median=279,
        lower whisker=185,
        lower quartile=187.25,
        upper quartile=280.75,
        upper whisker=283
      },
      ] coordinates {};
      \addplot+[
      boxplot prepared={
        median=270,
        lower whisker=180,
        lower quartile=182,
        upper quartile=271,
        upper whisker=272
      },
      ] coordinates {};
      \addplot+[
      boxplot prepared={
        median=271,
        lower whisker=181,
        lower quartile=181,
        upper quartile=270,
        upper whisker=273
      },
      ] coordinates {};

      \addplot+[
      boxplot prepared={
        median=132,
        lower whisker=130,
        lower quartile=131,
        upper quartile=171,
        upper whisker=174
      },
      ] coordinates {};
      \addplot+[
      boxplot prepared={
        median=187,
        lower whisker=185,
        lower quartile=186,
        upper quartile=279.25,
        upper whisker=282
      },
      ] coordinates {};
      \addplot+[
      boxplot prepared={
        median=182,
        lower whisker=180,
        lower quartile=181,
        upper quartile=271,
        upper whisker=273
      },
      ] coordinates {};
      \addplot+[
      boxplot prepared={
        median=181,
        lower whisker=179,
        lower quartile=182,
        upper quartile=271,
        upper whisker=272
      },
      ] coordinates {};

      \addplot+[
      boxplot prepared={
        median=131,
        lower whisker=129,
        lower quartile=130,
        upper quartile=131,
        upper whisker=134
      },
      ] coordinates {};      
      \addplot+[
      boxplot prepared={
        median=186,
        lower whisker=184,
        lower quartile=186,
        upper quartile=187,
        upper whisker=188
      },
      ] coordinates {};
      \addplot+[
      boxplot prepared={
        median=181,
        lower whisker=180,
        lower quartile=180,
        upper quartile=182,
        upper whisker=182
      },
      ] coordinates {};
      \addplot+[
      boxplot prepared={
        median=182,
        lower whisker=181,
        lower quartile=182,
        upper quartile=183,
        upper whisker=183
      },
      ] coordinates {};
    \end{axis}
  \end{tikzpicture}
        \vspace{-2mm}
        \caption{$k=50$}
        \label{fig:memory50}
    \end{subfigure}
    \begin{subfigure}{0.33\textwidth}
      \centering
        \pgfmathdeclarefunction{fpumod}{2}{%
    \pgfmathfloatdivide{#1}{#2}%
    \pgfmathfloatint{\pgfmathresult}%
    \pgfmathfloatmultiply{\pgfmathresult}{#2}%
    \pgfmathfloatsubtract{#1}{\pgfmathresult}%
    \pgfmathfloatifapproxequalrel{\pgfmathresult}{#2}{\def\pgfmathresult{5}}{}%
}
\begin{tikzpicture}
    \begin{axis}
    [width=\textwidth,
      cycle list={{store_color},{assign_color, pattern=custom north east lines, pattern color=assign_color},{share_color, pattern=crosshatchh, pattern color=share_color},{access_color, pattern=custom north west lines, pattern color=access_color}},
  legend pos = north west,
  legend style={legend cell align=left, align=left, fill=none, draw=none,inner sep=-0pt, row sep=0pt,at={(0,1.2)}},
  legend style={/tikz/every even column/.append style={column sep=1cm}},
      boxplot/box extend=0.10,
        boxplot={
      %
      %
      draw position={1/5 + floor(\plotnumofactualtype/4) + 1/5*fpumod(\plotnumofactualtype,4)},
  },
  xtick={0,1,2,...,60},
  x tick label as interval,
  xticklabels={%
      {10},%
      {20},%
      {30},%
      {40},%
      {50},%
  },
  x tick label style={
      align=center
  },
    boxplot/draw direction=y,
    ylabel={Memory (MB)},
    ylabel style={font=\scriptsize},
    xlabel={$n$},
    minor y tick num=4,
    yticklabel style={font=\scriptsize},
    xticklabel style={font=\scriptsize},
    xlabel style={font=\scriptsize},
    ylabel style={font=\scriptsize},
    scaled y ticks = base 10:-2,
        ylabel shift={-.5em},
            ymax = 300,
        ymin = 100,
        ytick distance=100,
    ]
      \addplot+[
      boxplot prepared={
        median=171.25,
        lower whisker=168,
        lower quartile=170.25,
        upper quartile=172.375,
        upper whisker=175
      },
      ]
      coordinates {};
      \addplot+[
      boxplot prepared={
        median=280,
        lower whisker=277,
        lower quartile=279,
        upper quartile=281,
        upper whisker=282
      },
      ] coordinates {};
      \addplot+[
      boxplot prepared={
        median=271,
        lower whisker=269,
        lower quartile=270,
        upper quartile=271.75,
        upper whisker=272
      },
      ] coordinates {};
      \addplot+[
      boxplot prepared={
        median=270,
        lower whisker=269,
        lower quartile=270,
        upper quartile=271,
        upper whisker=271
      },
      ] coordinates {};
      \addplot+[
      boxplot prepared={
        median=171,
        lower whisker=170,
        lower quartile=171,
        upper quartile=172,
        upper whisker=173
      },
      ] coordinates {};
      \addplot+[
      boxplot prepared={
        median=280,
        lower whisker=277,
        lower quartile=279,
        upper quartile=281,
        upper whisker=283
      },
      ] coordinates {};
      \addplot+[
      boxplot prepared={
        median=271,
        lower whisker=269,
        lower quartile=270,
        upper quartile=271,
        upper whisker=272
      },
      ] coordinates {};
      \addplot+[
      boxplot prepared={
        median=271,
        lower whisker=269,
        lower quartile=270,
        upper quartile=271.75,
        upper whisker=272
      },
      ] coordinates {};
      \addplot+[
      boxplot prepared={
        median=171,
        lower whisker=169,
        lower quartile=171,
        upper quartile=172,
        upper whisker=174
      },
      ] coordinates {};      
      \addplot+[
      boxplot prepared={
        median=280,
        lower whisker=277,
        lower quartile=279,
        upper quartile=281,
        upper whisker=283
      },
      ] coordinates {};
      \addplot+[
      boxplot prepared={
        median=271,
        lower whisker=269,
        lower quartile=270,
        upper quartile=271,
        upper whisker=272
      },
      ] coordinates {};
      \addplot+[
      boxplot prepared={
        median=271,
        lower whisker=269,
        lower quartile=270,
        upper quartile=271.75,
        upper whisker=272
      },
      ] coordinates {};
      \addplot+[
      boxplot prepared={
        median=171,
        lower whisker=129,
        lower quartile=169.75,
        upper quartile=172.25,
        upper whisker=174
      },
      ] coordinates {};
      \addplot+[
      boxplot prepared={
        median=280,
        lower whisker=185,
        lower quartile=278.75,
        upper quartile=281,
        upper whisker=282
      },
      ] coordinates {};
      \addplot+[
      boxplot prepared={
        median=271,
        lower whisker=180,
        lower quartile=269.75,
        upper quartile=271,
        upper whisker=273
      },
      ] coordinates {};
      \addplot+[
      boxplot prepared={
        median=270,
        lower whisker=179,
        lower quartile=269,
        upper quartile=271,
        upper whisker=273
      },
      ] coordinates {};

      \addplot+[
      boxplot prepared={
        median=134,
        lower whisker=129,
        lower quartile=131,
        upper quartile=171,
        upper whisker=173
      },
      ] coordinates {};      
      \addplot+[
      boxplot prepared={
        median=277,
        lower whisker=184,
        lower quartile=186,
        upper quartile=280.5,
        upper whisker=282
      },
      ] coordinates {};
      \addplot+[
      boxplot prepared={
        median=269,
        lower whisker=179,
        lower quartile=181,
        upper quartile=271,
        upper whisker=272
      },
      ] coordinates {};
      \addplot+[
      boxplot prepared={
        median=270,
        lower whisker=180,
        lower quartile=182,
        upper quartile=271,
        upper whisker=272
      },
      ] coordinates {};
    \end{axis}
  \end{tikzpicture}
         \vspace{-2mm}
        \caption{$k=75$}
         \label{fig:memory75}
    \end{subfigure}%
\caption{Memory usage distribution across blockchain nodes.}
	\label{fig:memory}
    \vspace{-6mm}
\end{figure*}

\subsection{Configuration}
\PP{Hardware} We used  50 AWS EC2 \texttt{t3.medium} instances to deploy a Hyperledger Fabric blockchain network. 
Each instance had 4GB vRAM with a 2-core 3.1 GHz CPU (Intel Xeon) and network bandwidth of 4.3 Gbps. 
Nodes were deployed in the \texttt{us-east-1} region. 

\PP{Parameter Setting} 
We used standard parameters to achieve 128-bit security.
We used standard Poseidon Hash parameters for the Merkle tree including width $t=6$, full rounds $R_F=8$ and partial rounds $R_p=130$ \cite{poseidon}.
We used MiMC hash function with rate $512$ and capacity $513$.
We implemented the Schnorr signature scheme to produce a 512-bit signature. 
For the ECIES encryption scheme, our implementation used a 256-bit public and secret key. 
We used Ed25519 curve of order $2^{255}-19$ for zero-knowledge proofs based on Bulletproofs, ECIES, and Schnorr. For Hyperledger Fabric, we set block size as 10 and the number of transaction confirmations as 1.

%
%
%
%
%
 
\subsection{Micro-Benchmark}

\autoref{tab:micro_bench} presents the processing overhead in terms of circuit complexity, processing latency, and memory usage as well as the bandwidth overhead of each operation.


\begin{table}
    \centering
        \caption{Performance of each data operation.}   \label{tab:micro_bench}
   \resizebox{0.5\textwidth}{!}{
    \begin{tabular}{|c|c|c|c|c|c|c|c|}
    \hline
    {\textbf{Data}} & \multicolumn{2}{c|}{\textbf{Circuit}} & \multicolumn{2}{c|}{\textbf{Delay (s)}} & \multicolumn{2}{c|}{\textbf{Mem (MB)}} & \textbf{B/W}  \\ \cline{2-7}
    \textbf{Op.} & \textbf{$\lvert C \rvert$} & \textbf{$\lvert M \rvert$} & $\zkProve$ & $\zkVerify$  & \textbf{$\zkProve$} & \textbf{$\zkVerify$} & \textbf{(KB)} \\ \hline
    {Store}  & 2,389  & 644 & 1.8  & 1.5   & 40.1 & 39.3 & 1.0 \\ 
    {Assign} & 87,315 & 37,640 & 23.2 & 8.6 & 165.6 & 94.7 & 1.4 \\ 
    {Share}  & 86,067  & 37,014 & 23.0 & 8.5 & 159.3 & 89.8 & 1.4 \\
    {Access} & 86,068 & 37,014 & 23.1 & 8.5 & 159.4 & 89.8 & 1.4 \\ \hline
    \end{tabular}
     }
     		
		\begin{tablenotes}[flushleft]
\small			\item $ \bullet $ $\lvert C \rvert$: \# constraints, $\lvert M \rvert$: \# multiplication gates.
		\end{tablenotes}
\vspace{-6mm}
\end{table}

%
%

\PP{Circuit Complexity}
Recording store operations incur the least circuit complexity, compared with other operations such as assign, share, and access.
This is because performing a store operation only invokes one $\PedG$ and one $\HshG$, while other operations invokes $\MemG$ along with other gadgets.
The constraint system of $\MemG$ is complicated and requires proving and verifying (in zero-knowledge) $d=64$ Poseidon hash pre-images, each of which incurs $ 3 \times t \times R_F + 3 \times R_p = 534 $ number of multiplication constraints \cite{poseidon}.
Assign operation incurs a little higher circuit complexity than share and access because of an extra $\HshG$ used for $\sn$.
Share has one constraint over access as it requires an additional constraint to calculate difference between the current timestamp, $\curr$ and the share expiry timestamp, $\ts$ (see \autoref{sec.access}).

\PP{Processing Latency} 
The processing latency depends on circuit complexity, especially the number of multiplication gates ($|M|$).
Store operation incurs a small latency compared with other operations because it does not require checking set membership, which attributes to large circuit complexity.
We include the Common Reference String (CRS) loading delay that is needed for the backend zero-knowledge proof (i.e., Bulletproofs), which takes around $1.5$ seconds (s) in total.
Without CRS, the proving and verification time for store operation is 0.35 s and 85 milliseconds (ms), respectively, which are about 60 times lesser than assign, share, and access.
This matches their difference in $\lvert M \rvert$.

\PP{Memory Usage}
The memory usage also depends on the circuit complexity to prove each data operation. 
The CRS component takes approximately $ 32 $ MB.
Proving incurs more memory usage than verification, due to the space needed to store the witness.
%






%

%
%
%
%

%
%

\PP{Bandwidth}
Since Bulletproof incurs a polylogarithmic proof size, the bandwidth overhead for each data operation is small.
The proof size produced by Bulletproofs is $2\lceil \log_2(|M|) \rceil +13$ group elements.
For store, $|M| = 644$ so its proof contains $ 2 \lceil \log_2(644) \rceil + 13 = 33$ group elements, which results in a total of $ 32 \cdot 33  = 1,056$ bytes (B). 
For assign, share, and access operations, $ |M|\approx 37,000 $ and, their proof size is 1,440 B.
%

%
%
%
%
%
%
%
%
%

\subsection{Macro-Benchmark}
We benchmark the performance of \sys~using Hyperledger Fabric.
We vary the number of concurrent records $k$ and blockchain nodes $n$ to evaluate the scalability of our proposed technique.

%
%


%

\PP{Overall latency}
\autoref{fig:exp:verificationlatency} presents the latency to verify a data operation with varying $k$ and $n$.
When $k=25$ and $n=10$, it takes around 17 s to verify a record of assign, share, and access operations (see \autoref{fig:latency_tx_fixed_25}).
We see that the verification delay reduces when increasing $n$. 
In particular, it reduces to around 11 s for $n=20$, and 9 s for $n=30$, after which, we observe no further improvement.
This is because, when $n=30$, every node only verifies one operation at a time. 
Notice that verifying an assign record should only take 8.5 s (from micro-benchmark), rather than 9 s as shown in \autoref{fig:exp:verificationlatency} (when $n \geq k$).
%
%
This is due to the delay caused by Hyperledger Fabric implementation of about 0.5 s.
The same trend is observed for store operation, where the latency is 4.13 s when $n=10$ and reduces to 2.51 s and 2 s when $n=20$ and $n=30$, respectively.

%

%
\autoref{fig:latency_tx_fixed_50} and \autoref{fig:latency_tx_fixed_75} shows the impact on verification latency when increasing $k$.
We observe that when doubling $k$,
the latency does not doubly increase but increases around 2 - 1.5 times.
This is due to the implementation of Hyperledger Fabric, 
which attempts to exploit all the CPU resources (2 cores in our setting) to verify all records submitted to it.
The speedup is not observed when $n=10$ because the load per node is 5 (when $k=50$) or 7.5 (when $k=75$). 
The 5/7.5 verifications happen at the same time across 2 CPU cores. 
This requires the CPUs to constantly switch between tasks which incurs a large delay.

\begin{figure}[t]
    \centering
    \begin{subfigure}[b]{0.4\textwidth}{
%
%
%
\begin{tikzpicture}
	
	\begin{axis}[
		cycle list={{store_color, mark=\storeshape}, {assign_color, mark=\assignshape}, {share_color, mark=\shareshape}, {access_color,mark=\accessshape}},
		width=.8\textwidth,
		height=.4\textwidth,
		at={(1.128in,0.894in)},
		scale only axis,		
		axis background/.style={fill=white},
		legend columns=4,
		ymajorgrids,
		xmajorgrids,
		grid style={line width=.5pt, draw=gray!90,dashed},
		major grid style={line width=.2pt,draw=gray!50},
		minor y tick num=5,
		axis background/.style={fill=white},	
		yticklabel style={font=\scriptsize},
		xticklabel style={font=\scriptsize},
		xlabel style={font=\scriptsize},
		ylabel style={font=\scriptsize},
		ylabel near ticks,
		ylabel shift=-5pt,
		yticklabel shift={0cm},
		xlabel={\# concurrent data operations ($k$)},
		xtick={0,1,2,3,4},
		xticklabels={1,25,50,75,100},
		ymin=0.0,
		ymax=60,
		ytick distance=15,
		ylabel={Time (sec)},
		ylabel style={font=\scriptsize},
		legend pos = north west,
		legend style={legend cell align=left, align=left, fill=none, draw=none,inner sep=-0pt, row sep=0pt,font=\tiny},
		legend columns=2, 
		]
		\addplot 
		table[row sep=crcr]{%
		0   4.05 \\
		1   4.2 \\
		2   4.75 \\
		3   5.99 \\
		4   7.5 \\
		};
		\addlegendentry{Store}
\addplot
table[row sep=crcr]{%
0   32.30 \\
1   32.36 \\
2   37.67 \\
3   41.37 \\
4   45.38 \\
};
\addlegendentry{Assign}
\addplot
table[row sep=crcr]{%
0   32.42 \\
1   32.23 \\
2   37.95 \\
3   40.95 \\
4   44.95 \\
};
\addlegendentry{Share}
\addplot
table[row sep=crcr]{%
0   31.1 \\
1   32.49 \\
2   36.89 \\
3   41.89 \\
4   46.48 \\
};
\addlegendentry{Access}
	\end{axis}

\end{tikzpicture}%
}
    \end{subfigure}
    \caption{End-to-end latency with $n=30$ blockchain nodes.}
    \label{fig:end_to_end_latency}
    \vspace{-6mm}
\end{figure}
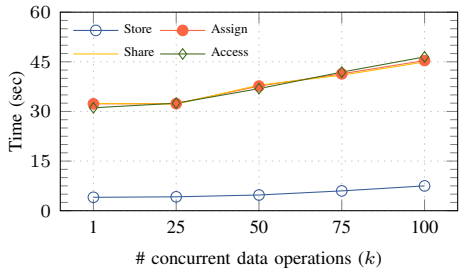

\autoref{fig:end_to_end_latency} presents end-to-end processing latency when $ n = 30 $ with varying $k$. 
It captures proving time (including overhead to load CRS), record creation time, verification time, and consensus delays. 

\PP{Memory Usage}
\autoref{fig:memory} presents distribution of memory used by nodes with varying $k$ and $n$.
In  \autoref{fig:memory25}, when $k=25$ and $n=10$, the median memory used to verify a store/assign/share/access operation is about 171/279/271/271 MB, respectively. 
Similar to latency, we see that the memory used by nodes to verify records of data operations reduces when increasing $n$.
For example, the inter-quartile range of memory usage when $n=20$ and $k=25$ for a store operation is 130-151 MB with a median of 132 MB.
This is because the majority of the nodes verify one store operation and some handle two as seen by the maximum value of 172 MB.
When $n=50$ and $k=25$, half the nodes verify one store operation and the other half are idle, therefore, we observe an inter-quartile range of 92-131 MB with a median of 110 MB.
The lower quartile of 92 MB shows the memory needed to run Hyperledger Fabric.
The upper quartile of 131 MB shows the memory needed to verify one store operation.
%
%
%
%

As seen in \autoref{fig:memory50} and \autoref{fig:memory75}, increasing $k$ increases the memory usage by the blockchain nodes as observed by the tighter inter-quartile range.
However, the maximum memory used, regardless of $k$, is about 174/282/272/271 MB for store/assign/share/access operations . This is equivalent to verifying two operations at a time since every node only has two CPUs.

\subsection{Comparison with prior works}


\begin{table}[ht!]
	\caption{Comparison of \sys~over prior works.}
	  \centering	
	\begin{threeparttable}
		\footnotesize
		\begin{tabular}{|c|c|c|c|c|}
			\hline
			\multirow{2}{0.1\textwidth}{\centering \textbf{Scheme}} & \multicolumn{3}{c|}{\textbf{Data Operation Latency (s)}}& \multirow{2}{0.1\textwidth}{\centering \textbf{Public Verifiability}}  \\ 
\cline{2-4}
 &\textbf{Store} & \textbf{Share} & \textbf{Access} & \\ 
			 \hline
			 Ghostor \cite{ghostor} & 5.4 & 8.5 & 7.0& \no  \\
            \hline
			 Calypso \cite{calypso_vldb20}  & 15.3 & 15.3 & 18.1&  \no  \\
			 \hline
			 \sys  & 36.3*  & 32.4 & 31.1& \yes  \\ \hline			 
		\end{tabular}
		
		\begin{tablenotes}[flushleft]
			\item $^*$This combines the cost of ``store'' and ``assign'' operations.
			
		\end{tablenotes}
	\end{threeparttable}\label{e2ecomparison}
\vspace{-6mm}
\end{table}

We conservatively compare the performance of our scheme with some prior works related to our setting including Ghostor \cite{ghostor} and Calypso \cite{calypso_vldb20}.
\autoref{e2ecomparison} presents the end-to-end delays incurred to perform one data operation of our scheme and prior works. 
Ghostor incurs the least delay of an average of 7 seconds mostly due to the asymmetric encryption and Tor anonymous communication. 
Calypso incurs the second highest delay of around 16 seconds for each operation mainly as it relies on verifiable secret sharing. 
%
%
A data operation in $\sys~$ takes about 33 seconds mostly due to zero-knowledge proof of membership, which accounts for about 80\% of the delay. 
However, despite being slower than prior works, we note that our scheme permits public verifiability for data operations thereby, making the audit process transparent.  
In prior works, the log must be validated by a trusted auditor and requires all the data operators to participate in the audit process. %

 \label{ch:exp}
   	\section{Related Work} \label{sec.literature_review}


%
\PP{Centralized Audit Log} Traditional data sharing/access schemes record data operations directly on the server that stores the data \cite{smitha,mhd}.
Such records are needed for audit purposes, where a third-party auditor determines whether any unauthorized accesses have been made in the system \cite{tifs,sc}.
However, with the introduction of regulations like HIPAA a \textit{secure logging} mechanism \cite{hipaa} is needed. 
Secure logging primitives \cite{jsyst,chen} rely on a trusted author whose signature is considered the root of trust for integrity or requires analyzing the entirety of the audit log to detect tampering due to the privacy-preserving nature of the records. 
Alternatively, prior works \cite{sgx,ma} use secure hardware to achieve immutable and privacy-preserving records. 
Since the audit log is maintained by the centralized server, 
it is vulnerable to a single point of failure, 
where the corrupted server can compromise 
the confidentiality and integrity of the audit log.

\PP{Decentralized Audit Log} 
Several works have proposed a decentralized audit log to address the single point of failure vulnerabilities. 
These schemes were designed for medical data \cite{Medrec, MeDShare}, supply chain \cite{pp_sc_1, pp_sc_2}, and financial applications \cite{su, rach}. 
Some schemes permit tamper resistance and access control to data objects being logged  \cite{auditable_access,globecom}.
However, these approaches do not hide the user identities (e.g., sender/receiver).
Several works attempted to achieve either full anonymity \cite{ghostor} or partially anonymous \cite{calypso_vldb20}.
There are several schemes that permit the validity of audit log to be verified publicly \cite{droplet} or privately \cite{ghostor, calypso_vldb20}. 
Private validity verifiability requires every user in the system to participate and prove the integrity of their records during the audit phase. 
 \label{ch:lit_rev}
   	\section{Conclusion}
We proposed \sys, a privacy-preserving and immutable audit log scheme.
\sys~achieved immutability and resistance to single point of failure by leveraging blockchain, and achieved public verifiability of record validity using zero-knowledge proofs.
We fully implemented \sys~and evaluated its performance by using Hyperledger Fabric and Amazon EC2.

%

 \label{ch:conc}
   	\section*{Acknowledgement}
   	\noindent This research is supported by BitShares and Robert Bosch (as an unrestricted gift), and the Commonwealth Cyber Initiative (CCI), an investment in the advancement of cyber R\&D, innovation, and workforce development. For more information about CCI, visit \url{www.cyberinitiative.org}.
   	\small{
	\bibliographystyle{IEEEtran}
	\bibliography{references}

\begin{thebibliography}{10}
\providecommand{\url}[1]{#1}
\csname url@samestyle\endcsname
\providecommand{\newblock}{\relax}
\providecommand{\bibinfo}[2]{#2}
\providecommand{\BIBentrySTDinterwordspacing}{\spaceskip=0pt\relax}
\providecommand{\BIBentryALTinterwordstretchfactor}{4}
\providecommand{\BIBentryALTinterwordspacing}{\spaceskip=\fontdimen2\font plus
\BIBentryALTinterwordstretchfactor\fontdimen3\font minus
  \fontdimen4\font\relax}
\providecommand{\BIBforeignlanguage}[2]{{%
\expandafter\ifx\csname l@#1\endcsname\relax
\typeout{** WARNING: IEEEtran.bst: No hyphenation pattern has been}%
\typeout{** loaded for the language `#1'. Using the pattern for}%
\typeout{** the default language instead.}%
\else
\language=\csname l@#1\endcsname
\fi
#2}}
\providecommand{\BIBdecl}{\relax}
\BIBdecl

\bibitem{tifs}
J.~Yuan and S.~Yu, ``Public integrity auditing for dynamic data sharing with
  multiuser modification,'' \emph{IEEE TIFS}, vol.~10, no.~8, 2015.

\bibitem{Liu_Access}
X.~Liu, Z.~Wang, C.~Jin, F.~Li, and G.~Li, ``A blockchain-based medical data
  sharing and protection scheme,'' \emph{IEEE Access}, vol.~7, 2019.

\bibitem{Medrec}
H.~Guo, W.~Li, M.~Nejad, and C.-C. Shen, ``Access control for electronic health
  records with hybrid blockchain-edge architecture,'' in \emph{IEEE
  Blockchain}, 2019.

\bibitem{Baralla}
W.~F. Lau, D.~Y.~W. Liu, and M.~H. Au, ``Blockchain-based supply chain system
  for traceability, regulation and anti-counterfeiting,'' in \emph{IEEE
  Blockchain}, 2021.

\bibitem{agri_suuply}
A.~Shahid \emph{et~al.}, ``Blockchain-based agri-food supply chain: A complete
  solution,'' \emph{IEEE Access}, vol.~8, 2020.

\bibitem{poms}
K.~Toyoda, P.~T. Mathiopoulos, I.~Sasase, and T.~Ohtsuki, ``A novel
  blockchain-based product ownership management system (poms) for
  anti-counterfeits in the post supply chain,'' \emph{IEEE Access}, vol.~5,
  2017.

\bibitem{iot_supply}
Q.~Song, Y.~Chen, Y.~Zhong, K.~Lan, S.~Fong, and R.~Tang, ``A supply-chain
  system framework based on internet of things using blockchain technology,''
  \emph{ACM Trans. Internet Technol.}, vol.~21, no.~1, jan 2021.

\bibitem{MetadataLeakage:zhou2020mobilogleak}
R.~Zhou, M.~Hamdaqa, H.~Cai, and A.~Hamou-Lhadj, ``Mobilogleak: A preliminary
  study on data leakage caused by poor logging practices,'' in \emph{2020 IEEE
  27th International Conference on Software Analysis, Evolution and
  Reengineering (SANER)}.\hskip 1em plus 0.5em minus 0.4em\relax IEEE, 2020,
  pp. 577--581.

\bibitem{NYRB13:Alan-Rusbridger-Metadata}
A.~Rusbridger, ``The {Snowden} leaks and the public,'' in \emph{The New York
  Review of Books---NYR Daily, 2013}.

\bibitem{hipaa}
{United States Department of Health}, ``{The Health Insurance Portability and
  Accountability Act of 1996}.''

\bibitem{jsyst}
I.~Ray, K.~Belyaev, M.~Strizhov, D.~Mulamba, and M.~Rajaram, ``Secure logging
  as a service—delegating record management to the cloud,'' \emph{IEEE
  Systems Journal}, 2013.

\bibitem{ma}
D.~Ma and G.~Tsudik, ``A new approach to secure logging,'' \emph{ACM Trans.
  Storage}, vol.~5, no.~1, mar 2009.

\bibitem{MeDShare}
Q.~Xia, E.~B. Sifah, K.~O. Asamoah, J.~Gao, X.~Du, and M.~Guizani, ``Medshare:
  Trust-less medical data sharing among cloud service providers via
  blockchain,'' \emph{IEEE Access}, vol.~5, 2017.

\bibitem{BPDS_globecom2018}
J.~Liu, X.~Li, L.~Ye, H.~Zhang, X.~Du, and M.~Guizani, ``Bpds: A blockchain
  based privacy-preserving data sharing for electronic medical records,'' in
  \emph{IEEE GLOBECOM}, 2018.

\bibitem{pp_sc_1}
F.~M. Benčić, P.~Skočir, and I.~P. Žarko, ``Dl-tags: Dlt and smart tags for
  decentralized, privacy-preserving, and verifiable supply chain management,''
  \emph{IEEE Access}, 2019.

\bibitem{pp_sc_2}
M.~E. Maouchi, O.~Ersoy, and Z.~Erkin, ``Decouples: A decentralized, unlinkable
  and privacy-preserving traceability system for the supply chain,'' in
  \emph{ACM SAC}, 2019.

\bibitem{droplet}
H.~Shafagh, L.~Burkhalter, S.~Ratnasamy, and A.~Hithnawi, ``Droplet:
  Decentralized authorization and access control for encrypted data streams,''
  in \emph{{USENIX} Security}, Aug. 2020.

\bibitem{ghostor}
Y.~Hu, S.~Kumar, and R.~A. Popa, ``Ghostor: Toward a secure {Data-Sharing}
  system from decentralized trust,'' in \emph{USENIX NSDI}, feb 2020.

\bibitem{calypso_vldb20}
E.~Kokoris-Kogias, E.~C. Alp, L.~Gasser, P.~Jovanovic, E.~Syta, and B.~Ford,
  ``Calypso: Private data management for decentralized ledgers,'' \emph{Proc.
  VLDB Endow.}, vol.~14, no.~4, Dec. 2020.

\bibitem{bulletproof}
B.~Bünz, J.~Bootle, D.~Boneh, A.~Poelstra, P.~Wuille, and G.~Maxwell,
  ``Bulletproofs: Short proofs for confidential transactions and more,'' in
  \emph{IEEE S\&P}, 2018.

\bibitem{zerocash_sp14}
E.~Ben~Sasson \emph{et~al.}, ``Zerocash: Decentralized anonymous payments from
  bitcoin,'' in \emph{IEEE S\&P}, 2014.

\bibitem{schnorr}
C.~P. Schnorr, ``Efficient identification and signatures for smart cards,'' in
  \emph{Springer CRYPTO}, 1990.

\bibitem{mimc}
M.~Albrecht, L.~Grassi, C.~Rechberger, A.~Roy, and T.~Tiessen, ``Mimc:
  Efficient encryption and cryptographic hashing with minimal multiplicative
  complexity,'' in \emph{IACR ASIACRYPT}, 2016.

\bibitem{dhaes}
M.~B. Michel~Abdalla and P.~Rogaway, ``Dhaes: An encryption scheme based on the
  diffie-hellman problem,'' IACR Cryptology ePrint Archive, Report 1999/007,
  1999.

\bibitem{fabric}
E.~Androulaki \emph{et~al.}, ``Hyperledger fabric: A distributed operating
  system for permissioned blockchains,'' in \emph{ACM EuroSys}, 2018.

\bibitem{pedersen}
T.~P. Pedersen, ``Non-interactive and information-theoretic secure verifiable
  secret sharing,'' in \emph{CRYPTO}, 1992.

\bibitem{poseidon}
L.~Grassi, D.~Khovratovich, C.~Rechberger, A.~Roy, and M.~Schofnegger,
  ``Poseidon: A new hash function for {Zero-Knowledge} proof systems,'' in
  \emph{USENIX Security}, Aug. 2021.

\bibitem{boneh}
D.~Boneh, B.~B{\"u}nz, and B.~Fisch, ``Batching techniques for accumulators
  with applications to iops and stateless blockchains,'' in \emph{Springer
  CRYPTO}, 2019.

\bibitem{smitha}
S.~Sundareswaran, A.~Squicciarini, and D.~Lin, ``Ensuring distributed
  accountability for data sharing in the cloud,'' \emph{IEEE TDSC}, vol.~9,
  no.~4, 2012.

\bibitem{mhd}
Z.~Yang, W.~Wang, and Y.~Huang, ``Ensuring reliable logging for data
  accountability in untrusted cloud storage,'' in \emph{IEEE ICC}, 2017.

\bibitem{sc}
H.~Tian \emph{et~al.}, ``Public audit for operation behavior logs with error
  locating in cloud storage,'' \emph{Soft Computing}, vol.~23, no.~11, Jun
  2019.

\bibitem{chen}
Y.~Shen, T.~Lam, J.-C. Liu, and W.~Zhao, ``On the confidential auditing of
  distributed computing systems,'' in \emph{IEEE ICDCS}, 2004, pp. 600--607.

\bibitem{sgx}
V.~Karande, E.~Bauman, Z.~Lin, and L.~Khan, ``Sgx-record: Securing system logs
  with sgx,'' in \emph{ACM Asia CCS}, 2017.

\bibitem{su}
H.~Al-Shaibani, N.~Lasla, and M.~Abdallah, ``Consortium blockchain-based
  decentralized stock exchange platform,'' \emph{IEEE Access}, vol.~8, 2020.

\bibitem{rach}
A.~Biryukov, D.~Khovratovich, and S.~Tikhomirov, ``Findel: Secure derivative
  contracts for ethereum,'' in \emph{Springer FC}, 2017.

\bibitem{auditable_access}
D.~{Di Francesco Maesa}, P.~Mori, and L.~Ricci, ``A blockchain based approach
  for the definition of auditable access control systems,'' \emph{Computers \&
  Security}, 2019.

\bibitem{globecom}
O.~J.~A. Pinno, A.~R.~A. Gregio, and L.~C.~E. De~Bona, ``Controlchain:
  Blockchain as a central enabler for access control authorizations in the
  iot,'' in \emph{IEEE GLOBECOM}, 2017.

\end{thebibliography}
	}
\end{document}